\documentclass[longauth]{aa}  

\usepackage{graphicx}
\usepackage{txfonts}
\usepackage{hyperref}
\def\numberofresults{471}
\def\gaia{Gaia}
\def\emcee{\texttt{emcee}}
\def\apsis{Apsis}

\def\edr3{Gaia EDR3}
\def\gdr3{Gaia DR3}
\def\gdrFour{Gaia DR4}
\def\gspphot{GSP-Phot}
\def\gspspec{GSP-Spec}
\def\flame{FLAME}
\def\esphs{ESP-HS}
\providecommand{\mass}{\ensuremath{\mathcal{M}}}
\providecommand{\azero}{\ensuremath{A_0}\xspace}

\providecommand{\abp}{\ensuremath{A_\mathrm{BP}}\xspace}
\providecommand{\arp}{\ensuremath{A_\mathrm{RP}}\xspace}

\providecommand{\ebpminrp}{\ensuremath{E(G_{\rm BP} - G_{\rm RP})}\xspace}

\providecommand{\orcit}[1]{\protect\href{https://orcid.org/#1}{\protect\includegraphics[width=8pt]{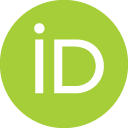}}}

\def\onlinedocucu8section{11}
\providecommand{\linktodoc}{https://geapre.esac.esa.int/archive/documentation/GDR3}
\providecommand{\linksec}[2]{\href{\linktodoc/Data_analysis/chap_cu8par/#1}{#2\xspace}}

\begin{document}

\title{Gaia Data Release 3: Analysis of the Gaia BP/RP spectra using the General Stellar Parameterizer from Photometry}
\titlerunning{Gaia DR3: Analysis of BP/RP spectra by \gspphot}
\authorrunning{R.~Andrae et al.}

\author{
             R.~                        Andrae\orcit{0000-0001-8006-6365}\inst{\ref{inst:0001}}\thanks{andrae@mpia-hd.mpg.de}
\and         M.~                     Fouesneau\orcit{0000-0001-9256-5516}\inst{\ref{inst:0001}}
\and         R.~                         Sordo\orcit{0000-0003-4979-0659}\inst{\ref{inst:0003}}
\and     C.A.L.~                  Bailer-Jones\inst{\ref{inst:0001}}
\and       T.E.~                 Dharmawardena\orcit{0000-0002-9583-5216}\inst{\ref{inst:0001}}
\and         J.~                       Rybizki\orcit{0000-0002-0993-6089}\inst{\ref{inst:0001}}
\and         F.~                     De Angeli\orcit{0000-0003-1879-0488}\inst{\ref{inst:0007}}
\and     H.E.P.~                  Lindstr{\o}m\inst{\ref{inst:0008},\ref{inst:0009},\ref{inst:0010}}
\and       D.J.~                      Marshall\orcit{0000-0003-3956-3524}\inst{\ref{inst:0011}}
\and         R.~                       Drimmel\orcit{0000-0002-1777-5502}\inst{\ref{inst:0008}}
\and       A.J.~                          Korn\orcit{0000-0002-3881-6756}\inst{\ref{inst:0013}}
\and         C.~                      Soubiran\orcit{0000-0003-3304-8134}\inst{\ref{inst:0014}}
\and         N.~                     Brouillet\orcit{0000-0002-3274-7024}\inst{\ref{inst:0014}}
\and         L.~                   Casamiquela\orcit{0000-0001-5238-8674}\inst{\ref{inst:0014},\ref{inst:0017}}
\and      H.-W.~                           Rix\orcit{0000-0003-4996-9069}\inst{\ref{inst:0001}}
\and         A.~                Abreu Aramburu\inst{\ref{inst:0019}}
\and       M.A.~                   \'{A}lvarez\orcit{0000-0002-6786-2620}\inst{\ref{inst:0020}}
\and         J.~                        Bakker\inst{\ref{inst:bakker}}
\and         I.~                Bellas-Velidis\inst{\ref{inst:0021}}
\and         A.~                       Bijaoui\inst{\ref{inst:0022}}
\and         E.~                    Brugaletta\orcit{0000-0003-2598-6737}\inst{\ref{inst:0023}}
\and         A.~                       Burlacu\inst{\ref{inst:0024}}
\and         R.~                      Carballo\orcit{0000-0001-7412-2498}\inst{\ref{inst:0025}}
\and         L.~                        Chaoul\inst{\ref{inst:0026}}
\and         A.~                     Chiavassa\orcit{0000-0003-3891-7554}\inst{\ref{inst:0022}}
\and         G.~                      Contursi\orcit{0000-0001-5370-1511}\inst{\ref{inst:0022}}
\and       W.J.~                        Cooper\orcit{0000-0003-3501-8967}\inst{\ref{inst:0029},\ref{inst:0008}}
\and       O.L.~                       Creevey\orcit{0000-0003-1853-6631}\inst{\ref{inst:0022}}
\and         C.~                       Dafonte\orcit{0000-0003-4693-7555}\inst{\ref{inst:0020}}
\and         A.~                    Dapergolas\inst{\ref{inst:0021}}
\and         P.~                    de Laverny\orcit{0000-0002-2817-4104}\inst{\ref{inst:0022}}
\and         L.~                    Delchambre\orcit{0000-0003-2559-408X}\inst{\ref{inst:0035}}
\and         C.~                      Demouchy\inst{\ref{inst:0036}}
\and         B.~                    Edvardsson\inst{\ref{inst:0037}}
\and         Y.~                    Fr\'{e}mat\orcit{0000-0002-4645-6017}\inst{\ref{inst:0038}}
\and         D.~                      Garabato\orcit{0000-0002-7133-6623}\inst{\ref{inst:0020}}
\and         P.~              Garc\'{i}a-Lario\orcit{0000-0003-4039-8212}\inst{\ref{inst:0040}}
\and         M.~             Garc\'{i}a-Torres\orcit{0000-0002-6867-7080}\inst{\ref{inst:0041}}
\and         A.~                         Gavel\orcit{0000-0002-2963-722X}\inst{\ref{inst:0013}}
\and         A.~                         Gomez\orcit{0000-0002-3796-3690}\inst{\ref{inst:0020}}
\and         I.~   Gonz\'{a}lez-Santamar\'{i}a\orcit{0000-0002-8537-9384}\inst{\ref{inst:0020}}
\and         D.~                Hatzidimitriou\orcit{0000-0002-5415-0464}\inst{\ref{inst:0045},\ref{inst:0021}}
\and         U.~                        Heiter\orcit{0000-0001-6825-1066}\inst{\ref{inst:0013}}
\and         A.~          Jean-Antoine Piccolo\orcit{0000-0001-8622-212X}\inst{\ref{inst:0026}}
\and         M.~                      Kontizas\orcit{0000-0001-7177-0158}\inst{\ref{inst:0045}}
\and         G.~                    Kordopatis\orcit{0000-0002-9035-3920}\inst{\ref{inst:0022}}
\and       A.C.~                     Lanzafame\orcit{0000-0002-2697-3607}\inst{\ref{inst:0023},\ref{inst:0052}}
\and         Y.~                      Lebreton\orcit{0000-0002-4834-2144}\inst{\ref{inst:0053},\ref{inst:0054}}
\and       E.L.~                        Licata\orcit{0000-0002-5203-0135}\inst{\ref{inst:0008}}
\and         E.~                       Livanou\orcit{0000-0003-0628-2347}\inst{\ref{inst:0045}}
\and         A.~                         Lobel\orcit{0000-0001-5030-019X}\inst{\ref{inst:0038}}
\and         A.~                         Lorca\inst{\ref{inst:0058}}
\and         A.~               Magdaleno Romeo\inst{\ref{inst:0024}}
\and         M.~                      Manteiga\orcit{0000-0002-7711-5581}\inst{\ref{inst:0060}}
\and         F.~                       Marocco\orcit{0000-0001-7519-1700}\inst{\ref{inst:0061}}
\and         N.~                          Mary\inst{\ref{inst:0062}}
\and         C.~                       Nicolas\inst{\ref{inst:0026}}
\and         C.~                     Ordenovic\inst{\ref{inst:0022}}
\and         F.~                       Pailler\orcit{0000-0002-4834-481X}\inst{\ref{inst:0026}}
\and       P.A.~                       Palicio\orcit{0000-0002-7432-8709}\inst{\ref{inst:0022}}
\and         L.~               Pallas-Quintela\orcit{0000-0001-9296-3100}\inst{\ref{inst:0020}}
\and         C.~                         Panem\inst{\ref{inst:0026}}
\and         B.~                        Pichon\orcit{0000 0000 0062 1449}\inst{\ref{inst:0022}}
\and         E.~                        Poggio\orcit{0000-0003-3793-8505}\inst{\ref{inst:0022},\ref{inst:0008}}
\and         A.~                  Recio-Blanco\orcit{0000-0002-6550-7377}\inst{\ref{inst:0022}}
\and         F.~                        Riclet\inst{\ref{inst:0026}}
\and         C.~                         Robin\inst{\ref{inst:0062}}
\and         R.~                 Santove\~{n}a\orcit{0000-0002-9257-2131}\inst{\ref{inst:0020}}
\and       L.M.~                         Sarro\orcit{0000-0002-5622-5191}\inst{\ref{inst:0076}}
\and       M.S.~                    Schultheis\orcit{0000-0002-6590-1657}\inst{\ref{inst:0022}}
\and         M.~                         Segol\inst{\ref{inst:0036}}
\and         A.~                       Silvelo\orcit{0000-0002-5126-6365}\inst{\ref{inst:0020}}
\and         I.~                        Slezak\inst{\ref{inst:0022}}
\and       R.L.~                         Smart\orcit{0000-0002-4424-4766}\inst{\ref{inst:0008}}
\and         M.~                  S\"{ u}veges\orcit{0000-0003-3017-5322}\inst{\ref{inst:0082}}
\and         F.~                  Th\'{e}venin\inst{\ref{inst:0022}}
\and         G.~                Torralba Elipe\orcit{0000-0001-8738-194X}\inst{\ref{inst:0020}}
\and         A.~                          Ulla\orcit{0000-0001-6424-5005}\inst{\ref{inst:0085}}
\and         E.~                       Utrilla\inst{\ref{inst:0058}}
\and         A.~                     Vallenari\orcit{0000-0003-0014-519X}\inst{\ref{inst:0003}}
\and         E.~                    van Dillen\inst{\ref{inst:0036}}
\and         H.~                          Zhao\orcit{0000-0003-2645-6869}\inst{\ref{inst:0022}}
\and         J.~                         Zorec\inst{\ref{inst:0089}}
}
\institute{
     Max Planck Institute for Astronomy, K\"{ o}nigstuhl 17, 69117 Heidelberg, Germany\relax                                                                                                                                                                                                                                                                       \label{inst:0001}
\and INAF - Osservatorio astronomico di Padova, Vicolo Osservatorio 5, 35122 Padova, Italy\relax                                                                                                                                                                                                                                                                   \label{inst:0003}\vfill
\and Institute of Astronomy, University of Cambridge, Madingley Road, Cambridge CB3 0HA, United Kingdom\relax                                                                                                                                                                                                                                                      \label{inst:0007}\vfill
\and INAF - Osservatorio Astrofisico di Torino, via Osservatorio 20, 10025 Pino Torinese (TO), Italy\relax                                                                                                                                                                                                                                                         \label{inst:0008}\vfill
\and Niels Bohr Institute, University of Copenhagen, Juliane Maries Vej 30, 2100 Copenhagen {\O}, Denmark\relax                                                                                                                                                                                                                                                    \label{inst:0009}\vfill
\and DXC Technology, Retortvej 8, 2500 Valby, Denmark\relax                                                                                                                                                                                                                                                                                                        \label{inst:0010}\vfill
\and IRAP, Universit\'{e} de Toulouse, CNRS, UPS, CNES, 9 Av. colonel Roche, BP 44346, 31028 Toulouse Cedex 4, France\relax                                                                                                                                                                                                                                        \label{inst:0011}\vfill
\and Observational Astrophysics, Division of Astronomy and Space Physics, Department of Physics and Astronomy, Uppsala University, Box 516, 751 20 Uppsala, Sweden\relax                                                                                                                                                                                           \label{inst:0013}\vfill
\and Laboratoire d'astrophysique de Bordeaux, Univ. Bordeaux, CNRS, B18N, all{\'e}e Geoffroy Saint-Hilaire, 33615 Pessac, France\relax                                                                                                                                                                                                                             \label{inst:0014}\vfill
\and GEPI, Observatoire de Paris, Universit\'{e} PSL, CNRS, 5 Place Jules Janssen, 92190 Meudon, France\relax                                                                                                                                                                                                                                                      \label{inst:0017}\vfill
\and ATG Europe for European Space Agency (ESA), Camino bajo del Castillo, s/n, Urbanizacion Villafranca del Castillo, Villanueva de la Ca\~{n}ada, 28692 Madrid, Spain\relax                                                                                                                                                                                      \label{inst:0019}\vfill
\and CIGUS CITIC - Department of Computer Science and Information Technologies, University of A Coru\~{n}a, Campus de Elvi\~{n}a s/n, A Coru\~{n}a, 15071, Spain\relax                                                                                                                                                                                             \label{inst:0020}\vfill
\and National Observatory of Athens, I. Metaxa and Vas. Pavlou, Palaia Penteli, 15236 Athens, Greece\relax                                                                                                                                                                                                                                                         \label{inst:0021}\vfill
\and Universit\'{e} C\^{o}te d'Azur, Observatoire de la C\^{o}te d'Azur, CNRS, Laboratoire Lagrange, Bd de l'Observatoire, CS 34229, 06304 Nice Cedex 4, France\relax                                                                                                                                                                                              \label{inst:0022}\vfill
\and INAF - Osservatorio Astrofisico di Catania, via S. Sofia 78, 95123 Catania, Italy\relax                                                                                                                                                                                                                                                                       \label{inst:0023}\vfill
\and Telespazio for CNES Centre Spatial de Toulouse, 18 avenue Edouard Belin, 31401 Toulouse Cedex 9, France\relax                                                                                                                                                                                                                                                 \label{inst:0024}\vfill
\and Dpto. de Matem\'{a}tica Aplicada y Ciencias de la Computaci\'{o}n, Univ. de Cantabria, ETS Ingenieros de Caminos, Canales y Puertos, Avda. de los Castros s/n, 39005 Santander, Spain\relax                                                                                                                                                                   \label{inst:0025}\vfill
\and CNES Centre Spatial de Toulouse, 18 avenue Edouard Belin, 31401 Toulouse Cedex 9, France\relax                                                                                                                                                                                                                                                                \label{inst:0026}\vfill
\and Centre for Astrophysics Research, University of Hertfordshire, College Lane, AL10 9AB, Hatfield, United Kingdom\relax                                                                                                                                                                                                                                         \label{inst:0029}\vfill
\and Institut d'Astrophysique et de G\'{e}ophysique, Universit\'{e} de Li\`{e}ge, 19c, All\'{e}e du 6 Ao\^{u}t, B-4000 Li\`{e}ge, Belgium\relax                                                                                                                                                                                                                    \label{inst:0035}\vfill
\and APAVE SUDEUROPE SAS for CNES Centre Spatial de Toulouse, 18 avenue Edouard Belin, 31401 Toulouse Cedex 9, France\relax                                                                                                                                                                                                                                        \label{inst:0036}\vfill
\and Theoretical Astrophysics, Division of Astronomy and Space Physics, Department of Physics and Astronomy, Uppsala University, Box 516, 751 20 Uppsala, Sweden\relax                                                                                                                                                                                             \label{inst:0037}\vfill
\and Royal Observatory of Belgium, Ringlaan 3, 1180 Brussels, Belgium\relax                                                                                                                                                                                                                                                                                        \label{inst:0038}\vfill
\and European Space Agency (ESA), European Space Astronomy Centre (ESAC), Camino bajo del Castillo, s/n, Urbanizacion Villafranca del Castillo, Villanueva de la Ca\~{n}ada, 28692 Madrid, Spain\relax                                                                                                                                                             \label{inst:0040}\vfill
\and Data Science and Big Data Lab, Pablo de Olavide University, 41013, Seville, Spain\relax                                                                                                                                                                                                                                                                       \label{inst:0041}\vfill
\and Department of Astrophysics, Astronomy and Mechanics, National and Kapodistrian University of Athens, Panepistimiopolis, Zografos, 15783 Athens, Greece\relax                                                                                                                                                                                                  \label{inst:0045}\vfill
\and Dipartimento di Fisica e Astronomia ""Ettore Majorana"", Universit\`{a} di Catania, Via S. Sofia 64, 95123 Catania, Italy\relax                                                                                                                                                                                                                               \label{inst:0052}\vfill
\and LESIA, Observatoire de Paris, Universit\'{e} PSL, CNRS, Sorbonne Universit\'{e}, Universit\'{e} de Paris, 5 Place Jules Janssen, 92190 Meudon, France\relax                                                                                                                                                                                                   \label{inst:0053}\vfill
\and Universit\'{e} Rennes, CNRS, IPR (Institut de Physique de Rennes) - UMR 6251, 35000 Rennes, France\relax                                                                                                                                                                                                                                                      \label{inst:0054}\vfill
\and Aurora Technology for European Space Agency (ESA), Camino bajo del Castillo, s/n, Urbanizacion Villafranca del Castillo, Villanueva de la Ca\~{n}ada, 28692 Madrid, Spain\relax                                                                                                                                                                               \label{inst:0058}\vfill
\and CIGUS CITIC, Department of Nautical Sciences and Marine Engineering, University of A Coru\~{n}a, Paseo de Ronda 51, 15071, A Coru\~{n}a, Spain\relax                                                                                                                                                                                                          \label{inst:0060}\vfill
\and IPAC, Mail Code 100-22, California Institute of Technology, 1200 E. California Blvd., Pasadena, CA 91125, USA\relax                                                                                                                                                                                                                                           \label{inst:0061}\vfill
\and Thales Services for CNES Centre Spatial de Toulouse, 18 avenue Edouard Belin, 31401 Toulouse Cedex 9, France\relax                                                                                                                                                                                                                                            \label{inst:0062}\vfill
\and Dpto. de Inteligencia Artificial, UNED, c/ Juan del Rosal 16, 28040 Madrid, Spain\relax                                                                                                                                                                                                                                                                       \label{inst:0076}\vfill
\and Institute of Global Health, University of Geneva\relax                                                                                                                                                                                                                                                                                                        \label{inst:0082}\vfill
\and Applied Physics Department, Universidade de Vigo, 36310 Vigo, Spain\relax                                                                                                                                                                                                                                                                                     \label{inst:0085}\vfill
\and Sorbonne Universit\'{e}, CNRS, UMR7095, Institut d'Astrophysique de Paris, 98bis bd. Arago, 75014 Paris, France\relax                                                                                                                                                                                                                                         \label{inst:0089}\vfill
\and European Space Agency (ESA), European Space Astronomy Centre (ESAC), Camino bajo del Castillo, s/n, Urbanizacion Villafranca del Castillo, Villanueva de la Ca\~{n}ada, 28692 Madrid, Spain\relax   
\label{inst:bakker}\vfill
}

\date{Received March 03, 2022; accepted May 04, 2022}

% \abstract{}{}{}{}{} 
% 5 {} token are mandatory
 
  \abstract
   {The astrophysical characterisation of sources is among the major new data products in the third Gaia data release (DR3). In particular, there are stellar parameters for \numberofresults~million sources estimated from low-resolution BP/RP spectra.}
   {We present the General Stellar Parameterizer from Photometry (\gspphot), which is part of the astrophysical parameters inference system (Apsis). \gspphot\  is designed to produce a homogeneous catalogue of parameters for hundreds of millions of single  non-variable stars based on their astrometry, photometry, and low-resolution BP/RP spectra. These parameters are effective temperature, surface gravity, metallicity, absolute $M_G$ magnitude, radius, distance, and extinction for each star.}
   {\gspphot\ uses a Bayesian forward-modelling approach to simultaneously fit the BP/RP spectrum, parallax, and apparent $G$ magnitude. A major design feature of \gspphot\ is the use of the apparent flux levels of BP/RP spectra to derive, in combination with isochrone models, tight observational constraints on radii and distances. We carefully validate the uncertainty estimates by exploiting repeat Gaia observations of the same source.}
   {The data release includes \gspphot\ results for \numberofresults~million sources with $G<19$.
   Typical differences to literature values are 110~K for $T_\textrm{eff}$ and 0.2-0.25 for $\log g$, but these depend strongly on data quality. In particular, \gspphot\ results are significantly better for stars with good parallax measurements ($\varpi/\sigma_\varpi>20$), mostly within 2kpc. Metallicity estimates exhibit substantial biases compared to literature values and are only useful at a qualitative level. However, we provide an empirical calibration of our metallicity estimates that largely removes these biases. Extinctions \azero\ and \abp\ show typical differences from reference values  of 0.07-0.09~mag. MCMC samples of the parameters are also available for 95\% of the sources.}
   {\gspphot\ provides a homogeneous catalogue of stellar parameters, distances, and extinctions that can be used for various purposes, such as sample selections (OB stars, red giants, solar analogues etc.). In the context of asteroseismology or ground-based interferometry, where targets are usually bright and have good parallax measurements, \gspphot\ results should be particularly useful for combined analysis or target selection.}


   \maketitle
%
%-------------------------------------------------------------------

\section{Introduction}

The ESA Gaia satellite \citep{2016A&A...595A...1G} observes nearly two billion sources, most of which are stars residing in our Milky Way galaxy. Its main objective is to measure the parallax and proper motions of  these stars with unprecedented accuracy. To achieve this goal, a correction dependent on the source colour is mandatory, and for this a low-resolution BP/RP spectrum is collected for each source. The DPAC Coordination Unit 8 with its astrophysical parameter inference system \citep[CU8 \apsis,][]{Apsis2013} classifies and determines the astrophysical parameters for these sources from the Gaia data. This allows more efficient exploitation of the exquisite astrometry and photometry offered by Gaia, for example\ by enabling appropriate selection criteria tailored to particular science cases. \gdr3 \citep{DR3-DPACP-185} will provide the first major release of results from CU8 \citep{DR3-DPACP-157,DR3-DPACP-160,DR3-DPACP-158}, including a general validation \citep{DR3-DPACP-127}.

In this paper, we describe the General Stellar Parameterizer from Photometry (\gspphot), which is one module in the CU8 \apsis\ chain described in \citet{Apsis2013}. \gspphot\ is designed to infer stellar parameters, distances, and line-of-sight extinctions from Gaia's low-resolution BP/RP spectra \citep{2021A&A...652A..86C,EDR3-DPACP-118}, astrometry \citep{2021A&A...649A...2L}, and photometry \citep{2021A&A...649A...3R}. In Gaia DR3, the Gaia archive provides \gspphot\ results for \numberofresults~million sources with apparent magnitude $G\leq 19$. We also draw attention to a second module of the CU8 \apsis\ chain, the General Stellar Parameterizer from Spectroscopy \citep[\gspspec;][]{DR3-DPACP-186}, which is also designed to characterise single stars in \gdr3 but using the higher resolution RVS spectra \citep{DR3-DPACP-154} instead of the low-resolution BP/RP spectra.

An early version of \gspphot\ was described in \citet{Liu2012} and the core methods were laid out in \citet{Ilium2010} and \citet{Aeneas2011}. Section~\ref{sect:new-features} provides an overview of the current version of \gspphot\ adopted for \gdr3\ and highlights the improvements over the earlier version in \citet{Liu2012}. Section~\ref{sect:application-to-DR3} then presents some scientific validation results from \gspphot\ when applied to \gdr3\ data. Further validation results from \gspphot\ are presented in \citet{DR3-DPACP-157} and \citet{DR3-DPACP-160}. We conclude in Sect.~\ref{sect:summary}.

\section{\gspphot\ in a nutshell}
\label{sect:new-features}

\subsection{Main principles}
\label{ssec:gspphot-main-principles}

The main goal of \gspphot\ is to characterise all single stars in the Gaia catalogue based on their astrometry, photometry and, most importantly, their low-resolution BP/RP spectra. Those data are available for most sources with $G<19$ in the Gaia catalogue. We emphasise that the BP/RP spectra are time-averaged mean spectra, which means that\ any intrinsic time variability is lost.
%(unlike, e.g., RVS spectra which are only available for sources brighter than $G\lesssim 16.2$). 
\gspphot\ aims to provide a homogeneously derived catalogue of stellar parameters for non-variable single stars for all Gaia sources for which BP/RP spectra are available (which includes sources whose BP/RP spectra are not published in \gdr3). Other modules in the \apsis\ chain treat stars in binary systems or specific subtypes of stars in more specialised ways \citep[see MSC and Extended Stellar Parametrizers in ][]{DR3-DPACP-157,Apsis2013}. We emphasise that \gspphot\ uses only Gaia data: one objective of \gspphot\ is to attach a consistent set of astrophysical labels to the Gaia data and to also show how well stars can be generically characterised from Gaia data alone. Moreover, using non-Gaia data would fold in systematic errors and selection effects from external catalogues, which would make it more difficult to trace issues back to data sets during validation.

\gspphot\ comprises one main algorithm whose results are published in \gdr3\ and two support algorithms whose results are used internally but are not published. The main algorithm is called Aeneas \citep[referred to as $q$-method in][]{Aeneas2011}; it fits the measured BP/RP spectra, parallax, and apparent $G$ magnitude (see Sect.~\ref{sect:predicting-observables}), thereby estimating the stellar parameters. For this optimisation process, Aeneas employs a specific type of Markov-chain Monte-Carlo (MCMC) sampling using an ensemble of walkers \citep{EMCEE2013}. More specifically, the ensemble MCMC optimises only four fit parameters, namely the stellar age, mass, metallicity (see Sect.~\ref{sect:forward-isochrones}), and the line-of-sight monochromatic extinction $A_0$ at 541.4~nm, where $A_0$ is the extinction parameter from the adopted Fitzpatrick extinction law \citep{1999PASP..111...63F};  \linksec{ssec:cu8par_inputdata_xp_SimuExtLaw}{see also Sect.~\onlinedocucu8section.2.3.1.4 in the online documentation for details}. Other parameters such as distance or the extinction $A_G$ in the $G$ band are derived (see Sect.~\ref{ssec:deriver-parameters}). As in \citet{Aeneas2011}, \gspphot\ invokes astrophysical prior information; for example\ a Hertzsprung--Russell diagram (see Sect.~\ref{sect:priors}).
The two support algorithms provide the initial guess for the MCMC: first, the machine-learning algorithm Extremely Randomised Trees \citep{extratrees2006} estimates stellar parameters directly from the BP/RP spectra; second, a gradient-descent algorithm \citep[Ilium,][]{Ilium2010} further improves this initial parameter estimate. This is necessary because the MCMC alone would require too much computation time to find the best parameters without such an initial guess (see Sect.~\ref{sect:computational-cost}).

Within its forward-modelling context, \gspphot\ results are tied to the choice of model SEDs used to create synthetic BP/RP spectra. In its current version, \gspphot\ uses four different sets of model SEDs covering different temperature ranges of stars (see Sect.~\ref{sect:multi-library-approach}). In Sect.~\ref{ssec:XP-model-mismatch}, we briefly investigate various different model SEDs and the extent to which their synthetic BP/RP spectra deviate from real observed BP/RP spectra.

\subsection{Predicting observables}
\label{sect:predicting-observables}

Combining multiple observables of different kinds is helpful to better constrain the model parameters and extract the maximum information out of all available measurements \citep[e.g.][]{Aeneas2011,Schoenrich2014}. Below, we outline the observable data that are available to constrain our model within the Gaia and \gspphot\ context. 

First and foremost, we have the low-resolution BP and RP spectra, which are available for most sources observed by Gaia \citep{EDR3-DPACP-118}. These are provided by CU5 in the format of coefficients for an adopted basis representation \citep{2021A&A...652A..86C}. \citet{EDR3-DPACP-120} estimate that the spectral resolution, $\frac{\lambda}{\Delta\lambda}$, of BP ranges from 20 to 60 and that of RP  from 30 to 50, where the higher resolution is achieved for shorted wavelengths for both BP and RP. For use in CU8, these continuous basis functions are then evaluated on a defined grid of physical wavelengths in order to produce actual sampled spectra in the common format of photon flux within a wavelength range (pixel). DPAC/CU5 also provide covariance matrices for the coefficients of BP and RP. As the CU8 wavelength sampling uses more pixels than coefficients that are provided by CU5, a pixel covariance matrix could be computed, but it would not have full rank and therefore could not be inverted to define a $\chi^2$. For \gdr3, CU8 only takes the diagonal elements of the pixel covariance matrix into account, but neglects the correlations. This approximation will be dropped in future versions of \gspphot.

Second, Gaia provides an apparent $G$ magnitude, which is available for all sources in the Gaia catalogue. The possibility to exploit the apparent $G$ magnitude was already envisaged in \citet{Aeneas2011} and \citet{Liu2012}. Nevertheless, no absolute magnitude was available from their chosen fit parameters, and so the information provided by the apparent $G$ magnitude could not be fully exploited. We resolve this limitation by invoking stellar isochrones as discussed in Sect.~\ref{sect:forward-isochrones}.\footnote{We cannot exploit the integrated $G_\textrm{BP}$ and $G_\textrm{RP}$ photometry because these do not provide independent measurements from the dispersed BP/RP spectra themselves. We could use the integrated $G_\textrm{RVS}$ magnitudes where available for bright sources, but while the RVS passband is provided in \citet{DR3-DPACP-155}, unfortunately this only became available  after our \gdr3\ processing.}

Finally, Gaia provides a parallax measurement for most of the sources. This can be used to constrain a distance estimate through an astrometric $\chi^2$ contribution to the total likelihood.

Each of these three observables (BP/RP spectra, apparent $G$, parallax) provides a $\chi^2$. These are summed to obtain a total $\chi^2$, that is,\ the \gspphot\ likelihood function is constrained by all three observables.

\subsection{Forward model based on isochrones}
\label{sect:forward-isochrones}

The key idea in \citet{Aeneas2011} was to take the apparent $G$ magnitude and make use of the flux conservation equation,
\begin{equation}\label{eq:distance-modulus-G}
    G = M_G+A_G+5\log_{10}(d) - 5
    \,\textrm{,}
\end{equation}
to allow information from the spectrum to constrain the distance. However, from the atmospheric parameters $T_\textrm{eff}$, $\log g$, and [M/H], it is not possible to uniquely assign an absolute magnitude. This is the well-known problem of inverse isochrone matching. Instead, \citet{Aeneas2011} and \citet{Liu2012} chose to adopt a Hertzsprung--Russell diagram as a prior distribution and marginalise over the unknown absolute $M_G$ magnitude.

In this version of \gspphot, we solve this problem by starting from fundamental stellar parameters, namely age, initial mass, and metallicity. Stellar isochrones then uniquely provide us with astrophysically self-consistent absolute $M_G$ magnitude, radius, effective temperature, and surface gravity for the given fundamental parameters (age, mass, [M/H]). The atmospheric parameters are then also used to compute a synthetic model spectrum through multilinear interpolation over a given grid of models (see Sect.~\ref{sect:multi-library-approach}). Given the absolute $M_G$ magnitude provided by the isochrone and the extinction parameter $A_0$, we can compute the extinction $A_G$ from the model SED, the extinction curve, and the $G$ passband, and use that to predict the observed apparent $G$ magnitude from Eq.~(\ref{eq:distance-modulus-G}). This prediction of the apparent $G$ magnitude, which has an observational error of a few millimagnitudes, provides a very tight constraint on our model parameters and benefits the estimation of the surface gravity in particular. More precisely, $G$ has measurement errors of a few milli-magnitudes; however, the main uncertainty is likely to be in the $G$ passband estimation used to make model predictions of $M_G$ and $A_G$. We therefore introduce an error floor of 0.05mag (see Eq.~(\ref{eq:chi2-apparent-G})) to also account for model errors that may stem from imperfect knowledge of the passband.

For the isochrone models, we adopt a grid of PARSEC 1.2S Colibri S37 models \citep[][and references therein]{2014MNRAS.445.4287T,2015MNRAS.452.1068C,2020MNRAS.498.3283P} with step sizes of 0.01 between 6.6 and 10.13 in logarithmic age (in years) and 0.03 between -4.15 and 0.80 in [M/H]. These very fine step sizes are required to allow for a computationally efficient 3D linear interpolation (see Sect.~\ref{sect:computational-cost}) over age, mass, and metallicity to obtain the derived parameters.

\subsection{Derived parameters}
\label{ssec:deriver-parameters}

The four (MCMC) fit parameters are logarithmic age, initial mass, metallicity [M/H], and the parameter $A_0$ in the extinction law \citep{1999PASP..111...63F}. Apart from the four fit parameters, there are several more derived parameters, though.

First, from the fit parameters, the isochrones provide us with derived values of effective temperature $T_\textrm{eff}$, surface gravity $\log g$, stellar radius $R$, and absolute $M_G$ magnitude. These additional parameters are derived within the astrophysical models underlying the isochrones themselves and are tabulated in the isochrone data.

Second, coupling the extinction $A_0$ and metallicity [M/H] together with $T_\textrm{eff}$ and $\log g$ from isochrones, we compute a model BP/RP spectrum from a library of models \citep[see Sect.~\ref{sect:multi-library-approach} and][]{DR3-DPACP-157}. This is done by computationally efficient 4D linear interpolation (see Sect.~\ref{sect:computational-cost}). We then use the fact that our model BP/RP spectra come with absolute flux levels that scale with $\sigma_B T_\textrm{eff}^4$ (Stefan–Boltzmann law). Hence, when we use such a model to fit an observed BP/RP spectrum, we obtain an analytic $\chi^2$ solution for the amplitude
\begin{equation}\label{eq:XP-amplitude}
a = \frac{R^2}{d^2}
    \,\textrm{,}
\end{equation}
which is needed to bring the model BP/RP spectrum to the flux scale of the observed BP/RP spectrum. Here, $R$ is the stellar radius and $d$ the distance of the star. As the radius is given by the isochrone, we can directly compute the distance $d$ from Eq.~(\ref{eq:XP-amplitude}) for every MCMC sample. This distance then also enters the likelihood by predicting the measured parallax,
\begin{equation}
\chi^2_\textrm{parallax} = \left(\frac{\varpi-\frac{1}{d}}{\sigma_\varpi}\right)^2
,\end{equation}
and observed apparent $G$ magnitude,
\begin{equation}\label{eq:chi2-apparent-G}
\chi^2_\textrm{G} = \frac{(G - M_G-A_G-5\log_{10}(d) + 5)^2}{\displaystyle 0.05^2 + \left(\frac{2.5\sigma_f}{f\log 10}\right)^2}
,\end{equation}
where $f$ and $\sigma_f$ are the measured apparent $G$ flux and its uncertainty and 0.05 acts as an error floor of 50 milli-magnitudes that is added in quadrature to the approximate magnitude error $\frac{2.5\sigma_f}{f\log 10}$ (propagated from flux $f$ and flux error $\sigma_f$). If $\chi^2_\textrm{spectra}$ denotes the chi-squared from fitting the observed BP/RP spectrum using the amplitude resulting from Eq.~(\ref{eq:XP-amplitude}), the combined log-likelihood is given by
\begin{equation}\label{eq:gspphot-likelihood-function}
\log\mathcal L = \textrm{const} -\frac{1}{2}\left(\chi^2_\textrm{spectra}+\chi^2_\textrm{parallax}+\chi^2_\textrm{G}\right)
,\end{equation}
ignoring irrelevant normalisation constants.
As explained in Sect.~\ref{ssec:filtering}, cases where the distance resulting from Eq.~(\ref{eq:XP-amplitude}) deviates too much from the measured parallax or is inconsistent with the apparent $G$ magnitude have been filtered out of \gdr3.

Third, we also need the extinction in the $G$ band, $A_G$, for Eq.~(\ref{eq:distance-modulus-G}). This is obtained from the SEDs underlying our grid of model BP/RP spectra. These SEDs cover the wavelength range from 300nm to 1100nm. We apply interstellar extinction to the model grid according to \citet{1999PASP..111...63F} assuming constant $R_0=3.1$ (\linksec{ssec:cu8par_inputdata_xp_SimuExtLaw}{see also Sect.~\onlinedocucu8section.2.3.1.4 in the online documentation}). These reddened SEDs are then integrated over the Gaia $G$ passband and the resulting magnitude can be compared to the corresponding value without extinction in order to obtain $A_G$. Thus, we can assign a value of $A_G$ to all models in our model grid. However, $A_G$ is not a free fit parameter. Instead, $A_G$ is submitted to the same 4D linear interpolation as the model BP/RP spectra themselves. In addition to $A_G$, we also compute extinction values \abp\ and \arp\ in exactly the same way. From those extinctions, we can compute the reddening $\ebpminrp=\abp-\arp$.

\subsection{Prior distributions}
\label{sect:priors}

The full posterior probability distribution sampled by \gspphot\ is given by Eq.~(\ref{eq:appendix-posterior}) as derived in Appendix~\ref{appendix:priors-derivation}. One might expect us to only put priors on the four fit parameters (age, initial mass, metallicity, and extinction). However, sometimes it is astrophysically more intuitive to impose priors on derived parameters; for example\ the Hertzsprung-Russell diagram on temperature and absolute magnitude. There are several prior factors in Eq.~(\ref{eq:appendix-posterior}), which we now explain.

First, the prior for $A_G$ is a delta distribution that fixes the value to the extinction obtained from integrating the SED. While this may not behave like a commonly seen prior distribution, it remains a prior distribution from a mathematical point of view. Likewise, the prior for radius is a delta distribution fixing $R$ to the value provided by the isochrone.

Second, the extinction is restricted to the range $A_0\in[0,10]$ and within this range we adopt an ad hoc extinction prior of exponential form, $P(A_0|d)\propto e^{-A_0/\mu}$, where the mean value $\mu$ depends on Galactic latitude $b$ and distance $d$,
\begin{equation}
\mu=\frac{1 + 9\sin b}{1000\cdot(1 + \exp\left[-(d - 100)/10\right])}
.\end{equation}
This specific functional form and the choice of coefficients is the result of several test runs, reducing the occurrence of spuriously large extinctions in the validation sample.

Third, the distance was restricted to the range from 1~pc to 100~kpc. Furthermore, we adopt a distance prior of the form $P(d)\propto d^2e^{-d/L}$ (as introduced in 
\citealt{2015PASP..127..994B}) where the length scale $L$ depends on Galactic coordinates and has been mapped from the Gaia Early Data Release 3 (\edr3) mock catalog of \citet{Rybizki2020} excluding the Large Magellanic Cloud (LMC) and the Small Magellanic Cloud (SMC). To this end, we binned the mock data in Galactic coordinates and computed the mean distance $\langle d\rangle$ in each bin which is an estimator of the length scale $L=\frac{1}{3}\langle d\rangle$ under the assumed prior distribution $P(d)\propto d^2e^{-d/L}$. The length scale is then interpolated over the grid in Galactic coordinates in order to provide a smooth distance--prior variation over the sky. However, we set the length scale of the prior to be a factor of ten smaller than the result from the \edr3\ mock catalogue in an attempt to suppress outliers with unreasonably large distances. This also improved the comparison of temperature estimates to literature values, for example. Unfortunately, our validation sample lacked sources at large distances and therefore failed to show that this leads to a systematic underestimation of distances by \gspphot\ (see Sect.~\ref{ssect:distances_vs_prior}). 

Fourth, Eq.~(\ref{eq:appendix-posterior}) contains the factor $P([M/H],T_\textrm{eff},\log g,M_G,\log_{10}\tau,\log_{10}\mass),$ where there is an inter-dependency between the six components. Following \citet{Aeneas2011}, we adopt a Hertzsprung-Russell-diagram prior, that is,~we approximate the last factor as $P([M/H],T_\textrm{eff},\log g,M_G)$. Our specific Hertzsprung-Russell-diagram prior was constructed from the Gaia Universe Model Snapshot \citep{2012A&A...543A.100R}. We note that the PARSEC isochrones used by \gspphot\ and the Hertzsprung-Russell-diagram prior derived from GUMS are not always consistent. This may cause discrepancies, for example for low-mass dwarfs \citep[see discussion of results on the Local Bubble in][]{DR3-DPACP-127}.

Finally, by adopting the forward isochrone modelling (see Sect~\ref{sect:forward-isochrones}), we restrict the parameters $T_\textrm{eff}$, $\log g$, $M_G$, and radius, which can only populate regions that are reached by isochrones. Although this is formally part of the likelihood function, using isochrones in this way introduces a significant amount of prior astrophysical information.

We note that some of the priors are on derived parameters (Sect.~\ref{ssec:deriver-parameters}) instead of fit parameters. This is somewhat uncommon but still formally correct in a Bayesian sense (see Appendix~\ref{appendix:priors-derivation}). Furthermore, we note that we mainly employ priors as regularisation in order to suppress spuriously large extinctions and distances. We could have used other priors that are more motivated by astrophysics, for example\ an initial mass function, but we find that such priors are too weak to compete with the likelihood and so cannot confine the parameters to plausible regions of the parameter space.

\subsection{Multi-library approach}
\label{sect:multi-library-approach}

\gspphot\ not only employs isochrone models but also requires model grids of synthetic spectral energy distributions (SEDs) of stellar atmosphere models. As introduced in \citet{Apsis2013}, \gspphot\ uses four different such libraries of synthetic SEDs: MARCS for $T_\textrm{eff}$ between 2500 and 8000~K, PHOENIX for $T_\textrm{eff}$ between 3000 and 10\,000~K, A-stars for $T_\textrm{eff}$ between 6000 and 20\,000~K, and OB for $T_\textrm{eff}$ between 15\,000 and 55\,000~K. More details about these model libraries are provided in \citet{DR3-DPACP-157}.

Results from each library are reported individually in \gdr3,\footnote{The results from individual libraries are provided in the Gaia archive table named \texttt{astrophysical\_parameters\_supp}.} in case users have preferences for one particular library. We do \emph{not combine} the different estimates. Instead, we recommend a best library for users who prefer a single result per star.\footnote{The best-library results are provided in the Gaia archive tables named \texttt{gaia\_source} and \texttt{astrophysical\_parameters}.}

We identify the best library from the log-posterior probabilities of the MCMC samples. Let $\theta$ denote the \gspphot\ parameters (temperature, extinction, distance, etc.) and $\vec x$ the BP/RP spectra, $\varpi$ the measured parallax, and $G$ the apparent magnitude. Then, $p(\theta_s|\vec x,\varpi,G)$ denotes the posterior probability of the $s$th MCMC sample where $s=1,2,\ldots,S$ and $S$ is the number of MCMC samples (same for all sources). We tested various different measures of goodness-of-fit in order to identify the best library, such as the maximum posterior value in the MCMC or the Bayesian evidence estimated by the harmonic mean \citep[e.g.][]{Wolpert2012}. In the end, we obtained the best results when identifying the best library as the one having the highest mean log-posterior value averaged over the MCMC samples:
\begin{equation}\label{eq:cu8par_apsis_gspphot_mean_log_posterior}
\langle \log p(\theta|\vec x,\varpi,G)\rangle=\frac{1}{S}\sum_{s=1}^S \log p(\theta_s|\vec x,\varpi,G)
.\end{equation}
We note that Eq.~(\ref{eq:cu8par_apsis_gspphot_mean_log_posterior}) corresponds to a Monte-Carlo estimate of the differential entropy,
\begin{equation}
h=-\int  p(\theta|\vec x,\varpi,G)\log p(\theta|\vec x,\varpi,G) \,d\theta
,\end{equation}
which means that\ $h\approx-\langle \log p(\theta|\vec x,\varpi,G)\rangle$. In other words, the best library is chosen to be the library whose posterior distribution has the lowest differential entropy, that is,\ it provides the most information about the source from the point of view of information theory. This identification scheme of the best library is adequate but not perfect (see Sect.~\ref{ssec:comparison-to-literature}).

\subsection{Computational cost}
\label{sect:computational-cost}

The objective of \gspphot\ is to provide stellar parameter estimates for hundreds of millions of stars in \gdr3\ (and ultimately all 1.8 billion sources expected in \gdrFour). However, given the Gaia data release planning, we only have a limited amount of time available for processing. In order to comply with these limited resources, \gspphot\ can only process sources with $G<19$ \citep[see][Table~1 therein]{DR3-DPACP-157}. The actual processing of \gspphot\ in producing the results for \gdr3\ took approximately 360\,000 CPU hours on about 1400 cores, which equates to\ 257 hours.

One consequence of limited computational resources is that we cannot afford long convergence phases in our MCMC sampling. We therefore need a good initial guess in order to accelerate convergence. This initial guess is provided in a two-step process, starting with a machine-learning algorithm called Extremely Randomised Trees \citep{extratrees2006}, which is one of the support algorithms within \gspphot\ mentioned in Sect.~\ref{ssec:gspphot-main-principles}.\footnote{Extremely Randomised Trees replace the Support Vector Regression previously used in \citet{Liu2012} because they are much easier to train.} As in \citet{Liu2012}, the resulting initial guess is then further refined by a Newton-Raphson algorithm \citep{Ilium2010}, which is the second support algorithm in \gspphot. These two previous steps are computationally inexpensive and allow us to get away with very short convergence and relaxation phases in our MCMC.

Furthermore, while \citet{Liu2012} employed a classic Metropolis-Hastings MCMC, we changed this to the \emcee\ \citep{EMCEE2013}. The reason is that it is impossible to configure the step size of the proposal distribution in the Metropolis-Hastings MCMC such that it works well when all sources have different signal-to-noise ratios. Conversely, the \emcee\ is an ensemble MCMC that does not require  any proposal distribution to be fine-tuned, and can provide very efficient sampling for all sources.  Appendix~\ref{appendix:mcmc-configuration} gives further details on the MCMC configuration.

Last but not least, a key choice is that \gspphot\ uses multilinear interpolation over rectangular model grids. Among all possible interpolation schemes, we found this to be computationally highly efficient.\footnote{There are smoothing algorithms that are computationally faster but make additional approximations, e.g. those~used in \citet{2022arXiv220103252F}.} First, there is a 3D linear interpolation over isochrones with a rectangular grid in age, [M/H] and initial mass. Second, there is a 4D linear interpolation over rectangular grids of $T_\textrm{eff}$, $\log g$, [M/H], and $A_0$. In contrast, \citet{Liu2012} used a thin-plate-spline smoothing, which has a computational cost that is about two to three orders of magnitude more expensive. Likewise, \gspphot\ cannot afford the computational cost of propagating interpolation errors using a Gaussian process as in \citet{Starfish2015}. This inevitably causes an underestimation of the uncertainties.

\subsection{Values, uncertainties, and MCMC chains provided in \gdr3}
\label{ssec:values-uncertainties-mcmc}

The parameter values and lower and upper confidence levels that are provided in \gdr3\ are the median and 16th and 84th percentiles of the MCMC samples, respectively. We choose the median value because the mean (or mode) values can lie outside the confidence interval, especially (but not exclusively) in the presence of outliers, and inferring the mode from the MCMC samples requires additional computational cost.

We emphasise that we only provide one-dimensional confidence levels, and not correlations between parameters. The reason is that a correlation matrix implies that the posterior probability is Gaussian, which is not the case. Instead, for most sources, the MCMC samples exhibit clear evidence of non-Gaussianity (e.g.~asymmetry, curved contours, heavy tails). We therefore provide MCMC samples themselves to enable the user to correctly propagate uncertainties through any subsequent analysis. For reasons of data volume, it is not possible to provide full MCMC chains for all sources. Therefore, MCMC chains are provided only for the best-library results. Even here, we provide a reduced sample comprising only the last 100 MCMC samples for most sources.\footnote{Note that the reported values and confidence levels are \textit{always} estimated from the full MCMC having 2000 samples, even if the reported MCMC is reduced to only 100 samples.} The full MCMC chain with all 2000 samples (see Appendix~\ref{appendix:mcmc-configuration}) is provided only for sources brighter than $G<12$ and for a random subset of 1\% of sources fainter than that. Moreover, due to the filtering described in Sect.~\ref{ssec:filtering}, around 10\% of best-library results have no MCMC samples, because the originally best library was filtered out together with its MCMC and another library stepped up to fill the role of best library. This can happen, for example,\ when the originally best library provided a good fit to the BP/RP spectra, with their numerous pixels dominating the likelihood function, but otherwise poorly predicted the parallax or apparent $G$ magnitude.

\section{Application to Gaia data}
\label{sect:application-to-DR3}

In this section, we show some basic validation results from \gspphot. We first discuss aspects of MCMC convergence and filtering of results. We then verify the results at the level of distributions, for example,~via a Hertzsprung-Russell diagram. Finally, we compare \gspphot\ results to literature values. Further complementary validation results are presented by \citet{DR3-DPACP-157}, \citet{DR3-DPACP-160}, \citet{DR3-DPACP-158}, \citet{DR3-DPACP-127},
\citet{DR3-DPACP-123}, and \citet{DR3-DPACP-144}.

\subsection{Mismatch between models and observed BP/RP spectra}
\label{ssec:XP-model-mismatch}

The forward modelling of BP/RP spectra by \gspphot\ relies heavily on the agreement between observed BP/RP spectra and models thereof. Unfortunately, this agreement is not perfect. In order to illustrate this, we take solar twins from \citet{2021MNRAS.504.1873Y} and select 18 twins with $A_0<0.001$mag. For each of these 18 solar twins, we rescale their observed BP/RP spectra to $G=15$ from their actual apparent $G$ magnitude in order to make their flux levels comparable to each other and to model spectra from PHOENIX and MARCS \citep[see][]{DR3-DPACP-157} as well as to the model spectrum \texttt{sun\_model\_001} from the CALSPEC library\footnote{\url{https://www.stsci.edu/hst/instrumentation/reference-data-for-calibration-and-tools/astronomical-catalogs/calspec}} \citep{1995AJ....110.1316B,2014PASP..126..711B,2020AJ....160...21B} and to the 3D Stagger model at $T_\textrm{eff}=5787$K, $\log g=4.44$, [Fe/H]$=$0 \citep{2010A&A...516A..13P}.\footnote{\url{http://npollux.lupm.univ-montp2.fr/DBPollux/PolluxAccesDB}}
\begin{figure}
\begin{center}
\includegraphics[width=\columnwidth]{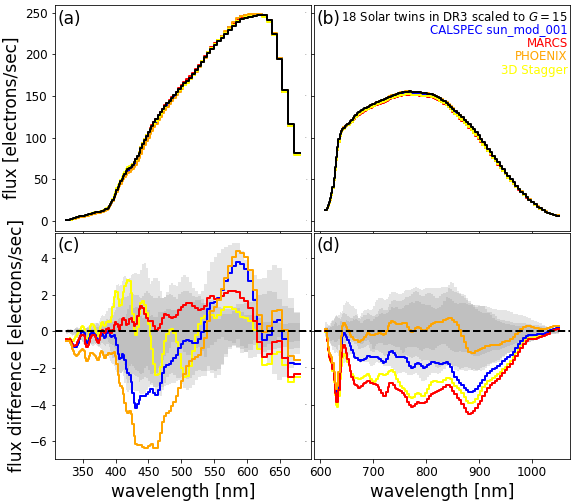}
\end{center}
\caption{Mismatch between observed BP/RP spectra of 18 solar twins from \citet{2021MNRAS.504.1873Y} with $A_0<0.001$mag (black lines and grey contours) and model BP/RP spectra from PHOENIX (orange lines), MARCS (red lines), CALSPEC \texttt{sun\_model\_001} (blue lines), and 3D Stagger (yellow lines). Panel (a): All BP spectra scaled to $G=15$. Panel (b): Same as panel (a) but for RP. Panel (c): Differences from median observed BP spectrum in panel (a). Panel (d): Same as panel (c) but for RP. Grey contours in panels (c) and (d) indicate the pixel-wise central 68\%, 90\%, and min/max intervals given the 18 solar twins.}
\label{fig:solar-twins-model-mismatch}
\end{figure}
In Fig.~\ref{fig:solar-twins-model-mismatch}a and b, we compare these models to the observed BP/RP spectra, which shows good agreement to first order. However, if we inspect the differences between models and observed BP/RP spectra in Fig.~\ref{fig:solar-twins-model-mismatch}c and d, we note that PHOENIX poorly matches BP, whereas MARCS poorly matches RP. CALSPEC \texttt{sun\_model\_001} is a poor match to both BP and RP. The 3D Stagger model matches BP reasonably well but not RP. The flux differences are as large as 10\% in BP and 4\% in RP per pixel, which are significant compared to the typical flux uncertainties of 2\% and below in BP and about 0.5\% in RP for sources in this apparent magnitude range, that is, between $G=7.4$ and $G=8.7$. We also emphasise that while we use Sun-like stars to illustrate this mismatch in Fig.~\ref{fig:solar-twins-model-mismatch}, it is very likely that also other types of stars suffer from similar mismatches.

This mismatch can only be partially ascribed to imperfections in the CU5 instrument model, namely where all models agree with each other but disagree with the observations \citep[in the steep RP cutoff 620-650nm and at the blue end of BP 320-400nm, see][]{EDR3-DPACP-120}. Nevertheless,  over wide
ranges, the various model spectra differ significantly  not only from the observed spectra but also from each other (400-650nm in BP, 680-950nm in RP). This mismatch can only originate from a genuine difference between the various model SEDs, as all models have their BP/RP spectra simulated with the exact same CU5 instrument model from \citet{EDR3-DPACP-120}. This systematic disagreement between models such as MARCS and PHOENIX most likely originates from different opacities which lead to different degrees of flux redistribution. Spectral lines are mostly invisible in low-resolution BP/RP spectra, such that the continuum shape is very important.

As a result of the systematic mismatch between models and observed BP/RP spectra, \gspphot\ results for parameters such as temperature and extinction often tend to cluster at the grid points of the model grids used for multilinear interpolation. This is visible as stripes when plotting \gspphot\ parameters. The reason is that the parameter optimisation struggles to make the model fit the observed BP/RP spectrum, which in the presence of systematic mismatches does not work perfectly. The closest fit can often only be achieved by letting the pixel fluxes of the model take a maximal or minimal value. As linear interpolation is a form of \textit{monotonic interpolation}, maxima or minima can only be acquired at grid points, but not in between grid points. In that sense, the presence of such stripes can be interpreted as an indicator of a mismatch between models and data.

\subsection{MCMC convergence}
\label{ssec:MCMC-convergence}

As explained in Sect.~\ref{sect:computational-cost} and Appendix~\ref{appendix:mcmc-configuration}, we have to use a fixed number of MCMC iterations in order to comply with the limited computational resources. This leads to the possibility of non-convergence. Visual inspection of 500 randomly chosen MCMC chains suggests that about 50\% of cases show at least minor evidence of non-convergence (e.g.\ drift in at least one parameter). However, those cases do not appear to correspond to outliers in scientific validation because test runs with longer MCMC chains and  better convergence did not yield better scientific results (e.g.\ in terms of lower differences to literature values). Instead, scientific outliers appear to be cases where the MCMC got stuck in a local optimum, that is,~converged to a bad solution.

\begin{figure}
\begin{center}
\includegraphics[width=\columnwidth]{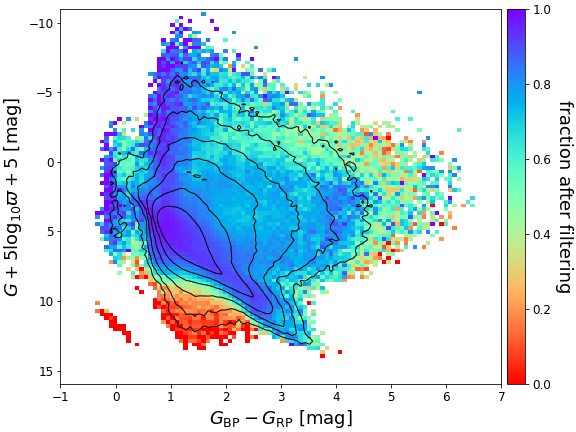}
\end{center}
\caption{Fraction of sources surviving the filtering described in Sect.~\ref{ssec:filtering} in the observed CMD. Contours indicate density of all sources with $G\leq 19$. This figure used a random subset of 2\,815\,418 sources drawn from the main catalogue. This plot does not include sources without parallax measurement or with negative parallax.}
\label{fig:CMD-impact-filtering}
\end{figure}

\subsection{Filtering of results}
\label{ssec:filtering}

\begin{figure}
\begin{center}
\includegraphics[width=\columnwidth]{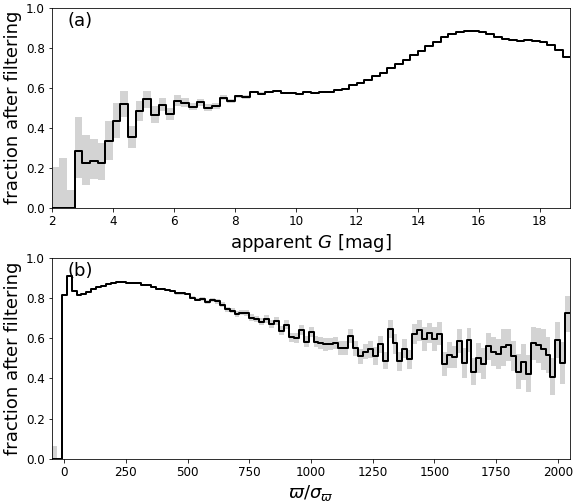}
\end{center}
\caption{Fraction of sources surviving the filtering described in Sect.~\ref{ssec:filtering} as a function of apparent $G$ magnitude (panel a) and parallax signal-to-noise ratio (panel b). Both panels use a random subset of about 100 million sources with $G<19$. Grey intervals indicate the horizontally decreasing intervals of 68\% confidence assuming a beta distribution in each bin.}
\label{fig:filtering-completeness-vs-Gmag}
\end{figure}

As mentioned in Sect.~\ref{sect:computational-cost} and further explained in \citet{DR3-DPACP-157}, \gspphot\ has only processed sources with $G<19$ due to limited computational resources.\footnote{Gaia sources with $G>19$ (about two-thirds of the entire Gaia sample)  have therefore never been processed with \gspphot. This is not due to a lack of data quality but simply due to a lack of computational resources.} However, while all sources with $G<19$ have been processed, not all the \gspphot\ results are published in \gdr3. Instead, based on initial validation work, we filter out results when one or more of the following conditions apply:
\begin{enumerate}[i]
\item There is no parallax available. In such a case, the distance estimate is usually unreliable and most of these sources would end up with improper solutions, for example\ in the colour--magnitude diagram.
\item The number of transits in the BP or RP spectrum is below 10 or 15, respectively. Such spectra are not of sufficient quality for the \gspphot\ analysis.
\item The observed apparent $G$ magnitude is poorly predicted, such that $M_G+A_G+5\log_{10}d-5$ differs from $G$ by more than 0.1mag.
\item The inverse distance differs from the measured parallax by more than ten times the parallax error.\label{filter-parallax-vs-inverse-distance}
\item The MCMC acceptance rate is below 10\%, suggesting that the initial guess was poor and did not allow the MCMC to properly explore the parameter space (see Sect.~\ref{ssec:MCMC-convergence}).
\end{enumerate}
Of the original 575.9 million sources with $G<19$, \numberofresults~million survive the filtering process ($\sim$81.7\%).\footnote{The vast majority of filtered cases for $G>12$ are due to a mismatch of observed and predicted apparent $G$ magnitudes.} Figure~\ref{fig:CMD-impact-filtering} shows where the sources lost due to filtering reside in the CMD. As \gspphot\ has no specific models for white dwarfs, almost all white dwarfs are lost. Otherwise, Fig.~\ref{fig:CMD-impact-filtering} shows that no other population is selectively affected by the filtering process. Furthermore, Fig.~\ref{fig:filtering-completeness-vs-Gmag}a shows that the completeness is around 0.8 for apparent $G$ magnitudes of between 16 and 18, whereas for magnitudes brighter than $G=12$ the completeness is around 0.5, falling below 0.5 for the very brightest sources at $G<5$.
Intuitively, one may expect higher completeness at the bright end due to better data quality. However, as is evident from Fig.~\ref{fig:filtering-completeness-vs-Gmag}b, the high parallax quality at the bright end is exactly where the \gspphot\ completeness is lowest. The main reason for this behaviour is the filter (\ref{filter-parallax-vs-inverse-distance}) requiring that the inverse \gspphot\ distance agrees to the parallax within ten times the parallax measurement error. As \gspphot\ infers the distance from the amplitude of the BP/RP spectra (see Sect.~\ref{ssec:deriver-parameters}), the noise on the BP/RP amplitude makes it increasingly difficult to agree with the parallax to within $10\sigma$ as the parallax measurement error decreases at the bright end.
%
%Likewise,  Fig.~\ref{fig:filtering-completeness-vs-Gmag}b shows that the completeness is above $\sim$80\% for most positive parallax signal-to-noise ratios but declines for sources with extremely high parallax quality, eventually reaching a plateau around 0.5 for sources with ($\frac{\varpi}{\sigma_\varpi}\gtrsim 1000$) which likely correspond to the bright $G<12$ regime from Fig.~\ref{fig:filtering-completeness-vs-Gmag}a. Yet, the completeness drops sharply for sources with negative parallaxes.

\subsection{CMD, HRD, and $T_\textrm{eff}$-$\log g$ diagrams}

\begin{figure*}[h!]
\begin{center}
\includegraphics[width=2\columnwidth]{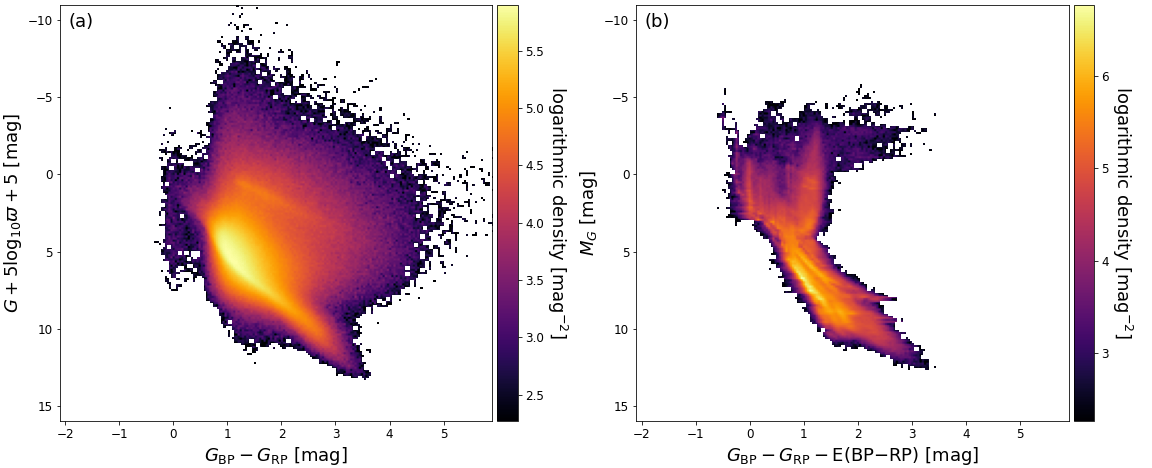}
\end{center}
\caption{Observed colour--magnitude diagram (panel a) and de-reddened colour--magnitude diagram (panel b). Both panels use the same sample of 2\,598\,519 stars that has been randomly selected from the main catalogue (see Appendix~\ref{appendix:example-ADQL-queries} for the ADQL query).}
\label{fig:CMD-de-reddening}
\end{figure*}

\begin{figure*}
\begin{center}
\includegraphics[width=2\columnwidth]{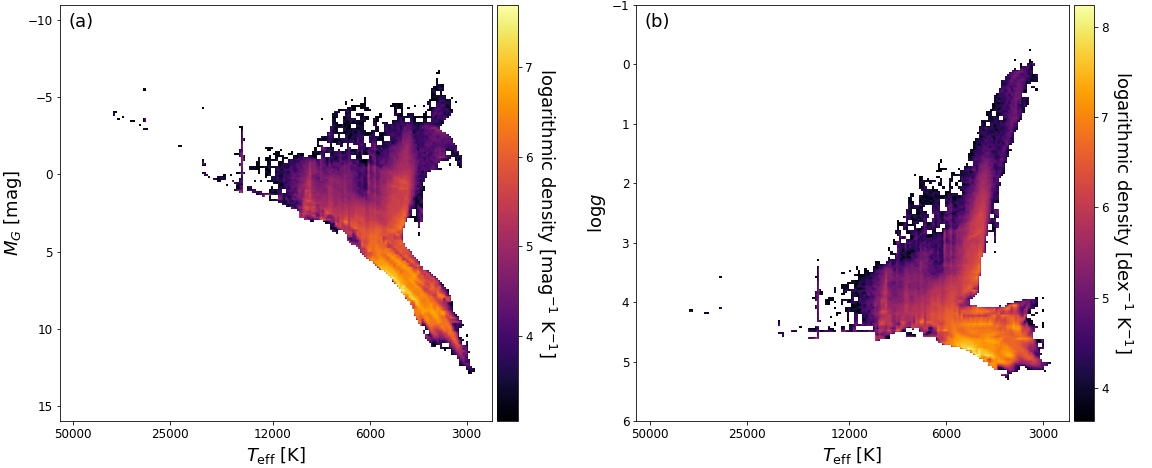}
\end{center}
\caption{Hertzsprung-Russell diagram (panel a) and $T_\textrm{eff}$-$\log g$ diagram (panel b). The same sample as in Fig.~\ref{fig:CMD-de-reddening} is used. The ADQL query for this plot can be found in Appendix~\ref{appendix:example-ADQL-queries}.}
\label{fig:HRD-and-Kiel}
\end{figure*}

The goal of \gspphot\ is to characterise all single stars in the Gaia catalogue. In Fig.~\ref{fig:CMD-de-reddening}, we demonstrate that the reddening and the absolute magnitude estimated by \gspphot\ indeed produce a de-reddened CMD that appears astrophysically plausible. Likewise, the Hertzsprung-Russell diagram and $T_\textrm{eff}$-$\log g$ diagram shown in Fig.~\ref{fig:HRD-and-Kiel} appear plausible. Nevertheless, we do see a prominent vertical stripe  at $T_\textrm{eff}=15\,000$K in both panels of Fig.~\ref{fig:HRD-and-Kiel}, which is a pile-up effect at the lower boundary of the OB library. We also see some minor vertical stripes at various temperatures, which are most likely a result of multilinear interpolation struggling in the presence of the mismatch between models and real BP/RP spectra (see discussion in Sect.~\ref{ssec:XP-model-mismatch}). The Hertzsprung-Russell diagram in Fig.~\ref{fig:HRD-and-Kiel}a also shows a hook at around 4000K protruding out of the giant population towards fainter magnitudes.

As the \gspphot\ forward model uses isochrones (see Sect.~\ref{sect:forward-isochrones}), we see in Fig.~\ref{fig:CMD-de-reddening}b and Fig.~\ref{fig:HRD-and-Kiel} that only the regions covered by these isochrones are populated, that is,\ there is no extrapolation.

\subsection{Comparison to literature values}
\label{ssec:comparison-to-literature}

We compare our parameter estimates to those from the literature, specifically 256\,967 stars from APOGEE DR16 \citep{ApogeeDR16}, 169\,825 stars from GALAH DR3 \citep{GalahDR3}, 513\,669 stars from LAMOST DR4 \citep{2011RAA....11..924W,2014IAUS..306..340W}, and 153\,284 stars from RAVE DR6 \citep{RaveDR6}. We note that these literature values were estimated from spectra with resolutions ranging from 1000 (LAMOST) to 28\,000 (GALAH). These spectral resolutions are significantly higher than those of the BP/RP spectra \citep[20-60 for BP, 30-50 for RP,][]{EDR3-DPACP-120}.

\begin{table}
\caption{Comparison of absolute differences between best $T_\textrm{eff}$ estimates (in K) from \gspphot\ and literature values. The columns indicate, from left to right, the median absolute difference (MedAD), the mean absolute difference (MAD), the root-mean-square difference (RMSD), the absolute difference not exceeded by 75\% of sources (AD 75\%), and the absolute difference not exceeded by 90\% of sources (AD 90\%).}
\label{table:teff-comparison-to-literature}
\centering
\begin{footnotesize}
\begin{tabular}{l|r|r|r|r|r}
catalogue & MedAD & MAD & RMSD & AD 75\% & AD 90\% \\
\hline
APOGEE & 169 & 418 & 1294 & 440 & 824 \\
GALAH & 110 & 150 & 228 & 198 & 315 \\
LAMOST & 110 & 156 & 253 & 198 & 327 \\
RAVE & 160 & 227 & 390 & 296 & 483 \\
\end{tabular}
\end{footnotesize}
\end{table}

\subsubsection{Effective temperature}

A major hurdle for \gspphot\ parameter estimates in general and effective temperatures in particular is the temperature--extinction degeneracy. This latter originates from the fact that a red star could be genuinely cool or have a higher temperature but is subject to notable line-of-sight dust attenuation and according reddening. Using only the low-resolution optical BP/RP spectra, it is very difficult to distinguish between these two cases. Employing the parallax and apparent magnitude can mitigate but not fully break this degeneracy \citep{Aeneas2011}. Nevertheless, the \gspphot\ temperatures are affected to some degree. Conversely, the effective temperatures from, for example, APOGEE, GALAH, LAMOST, or RAVE are derived from absorption lines in spectra of much higher resolution and are therefore unaffected by extinction.

Table~\ref{table:teff-comparison-to-literature} compares the absolute differences between our $T_\textrm{eff}$ estimates and literature values. We can see that there is only a mild dependence on the reference catalogue and that overall the median absolute difference shows that half of our results agree with literature values to within $\sim$170~K. For GALAH DR3 and LAMOST DR4 in particular, half of our results agree to within 110~K. The deviations are larger for APOGEE, which includes a significant fraction of red giants with relatively high extinction. For these stars, the temperature--extinction degeneracy is particularly difficult to break for \gspphot, resulting in hot stars being favoured and the OB library being labelled as best library. If we exclude all results from the OB library, the RMS difference reduces to 662 K, which is still higher than for the other catalogues.

\begin{figure}[h!]
\begin{center}
\includegraphics[width=\columnwidth]{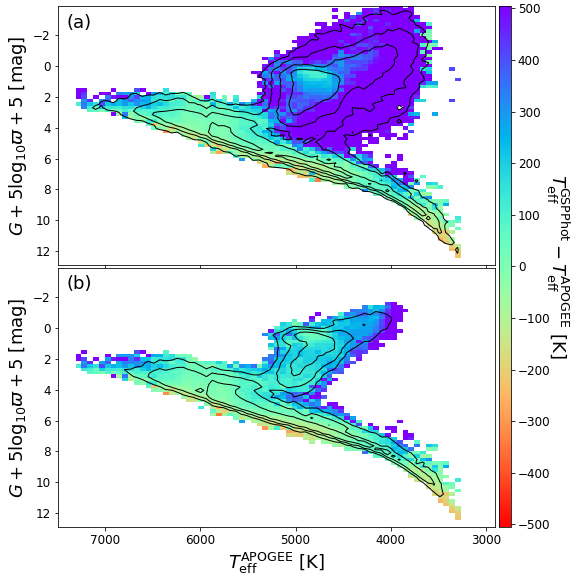}
\end{center}
\caption{Differences in $T_\textrm{eff}$ between results and literature values from APOGEE DR16 \citep{ApogeeDR16} in the Hertzsprung-Russell diagram. Panel (a): All sources. Panel (b): Sources with $\varpi/\sigma_\varpi>20$. Contours indicate source density decreasing by factors of 3.}
\label{fig:APOGEE-Teff-best-residuals-in-HRD}
\end{figure}

In order to exclude the possibility that APOGEE could have any internal inconsistency in itself, we also compare the results for the subset of 4015 stars shared by APOGEE DR16, GALAH DR3, and our results. On this specific subset, the RMS difference between \gspphot\ $T_\textrm{eff}$ estimates and APOGEE and GALAH values is 269K  and 263K, respectively, whereas the RMS difference between APOGEE and GALAH values is only 116K. This suggests that GALAH DR3 and APOGEE DR16 are in good mutual agreement and that we get a genuine overestimation of $T_\textrm{eff}$ in \gspphot\ for stars with line-of-sight extinctions $A_0\gtrsim 2$. Figure~\ref{fig:APOGEE-Teff-best-residuals-in-HRD}a reveals that the largest temperature differences occur for red giant stars, whereas \gspphot\ estimates appear to be consistent for main sequence stars. However, for stars with high-quality parallaxes, the \gspphot\ temperature estimates are much better and still usable in the red giant branch, as is evident from Fig.~\ref{fig:APOGEE-Teff-best-residuals-in-HRD}b. If we impose a parallax quality cut of $\frac{\varpi}{\sigma_\varpi}>20$, the median absolute deviation and the RMS deviation drop from 169K and 1294K (Table~\ref{table:teff-comparison-to-literature}) to 105K and 369K, respectively, although the number of stars also decreases by nearly a factor of two. Therefore, this appears to be not only due to the temperature--extinction degeneracy, but also in part to the systematic underestimation of distances caused by an overly harsh distance prior \citep{DR3-DPACP-160}.

Considering the subset of 6560 stars shared by GALAH DR3, RAVE DR6, and our results, the RMS difference between \gspphot\ $T_\textrm{eff}$ estimates and those of GALAH and RAVE is 230~K and  281~K, respectively, whereas the RMS difference between GALAH and 
RAVE is 214~K. This suggests that, for this subset, the \gspphot\ results are fully consistent with the typical uncertainties in the literature values.

\begin{figure}[h!]
\begin{center}
\includegraphics[width=\columnwidth]{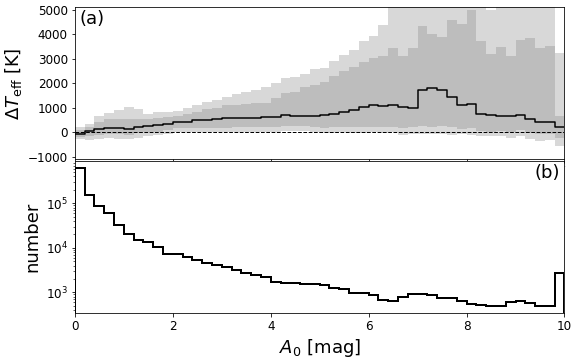}
\end{center}
\caption{Impact of extinction onto differences in $T_\textrm{eff}$ between results and literature values. Panel (a): Differences in $T_\textrm{eff}$ between our \gspphot\ results and literature values as function of estimated extinction $A_0$ for the joint samples from APOGEE DR16 \citep{ApogeeDR16}, GALAH DR3 \citep{GalahDR3}, LAMOST DR4 \citep{2011RAA....11..924W,2014IAUS..306..340W}, and RAVE DR6 \citep{RaveDR6}. If the same star appears in more than one literature resource, it enters this plot multiple times with identical \gspphot\ results but different literature values. The data are binned into $A_0$ ranges of 0.2mag, where the solid black line indicates the median residual and the two shaded regions show the 5th-to-95th percentiles and the 16th-to-84th percentiles, respectively. Panel (b): Distribution of $A_0$ estimates in this sample.}
\label{fig:Aeneas-Teff-residuals-vs-A0}
\end{figure}

Figure~\ref{fig:Aeneas-Teff-residuals-vs-A0}a reveals the temperature--extinction degeneracy very clearly. \gspphot\ overestimates $T_\textrm{eff}$ with rising $A_0$  estimate, because an overestimated extinction can compensate for an overestimated temperature.  This is hardly surprising given that \gspphot\ is using only optical data. However, for $A_0\lesssim 4$, the 84th percentile does not exceed 1000K, that is,~in that regime our temperature estimates are reasonably stable. Furthermore, Fig.~\ref{fig:Aeneas-Teff-residuals-vs-A0}b shows that \gspphot\ estimates a low extinction for the vast majority of stars. Here, we emphasise that, in principle, the temperature--extinction degeneracy works both ways: it can also cause a simultaneous underestimation of $T_\textrm{eff}$ and $A_0$. However, the majority of our validation targets with literature values usually have very low extinction.\footnote{This is a consequence of ground-based spectroscopic surveys preferentially targeting relatively bright sources.} If the true extinction is already close to zero, then there will be no room to significantly underestimate $A_0$  because of \gspphot's non-negativity constraint on $A_0$. In Sect.~\ref{ssec:teff-a0-degeneracy-randomly-split-sources}, we use a sample that is not restricted to low-extinction stars and it indeed shows the temperature--extinction degeneracy working both ways.

\begin{figure*}[h!]
\begin{center}
\includegraphics[width=2\columnwidth]{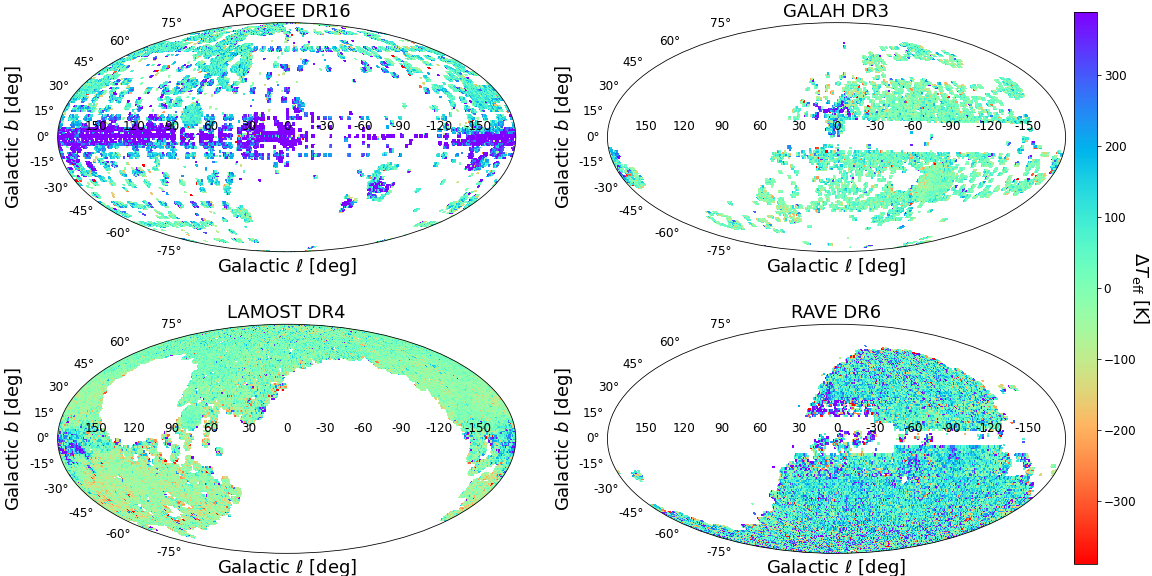}
\end{center}
\caption{Skymaps of differences between our results and literature $T_\textrm{eff}$ for APOGEE DR16 \citep{ApogeeDR16}, GALAH DR3 \citep{GalahDR3}, LAMOST DR4 \citep{2011RAA....11..924W,2014IAUS..306..340W}, and RAVE DR6 \citep{RaveDR6}. All skymaps use the Mollweide projection where lines of constant latitude are horizontal straight lines parallel to the equator.}
\label{fig:skymaps-Aeneas-Teff-residuals}
\end{figure*}

The skymaps of temperature differences in Fig.~\ref{fig:skymaps-Aeneas-Teff-residuals} also show that \gspphot\ typically overestimates $T_\textrm{eff}$ for APOGEE targets, whereas the results are typically much better  for targets from other catalogues. Nevertheless, all literature catalogues suggest that we tend to overestimate $T_\textrm{eff}$ in the Galactic plane. Again, this is a consequence of the temperature--extinction degeneracy.

\begin{table}
\caption{Comparison of best $\log g$ estimate from \gspphot\ to literature values. Columns as in Table~\ref{table:teff-comparison-to-literature}.}
\label{table:logg-comparison-to-literature}
\centering
\begin{footnotesize}
\begin{tabular}{l|r|r|r|r|r}
catalogue & MedAD & MAD & RMSD & AD 75\% & AD 90\% \\
\hline
APOGEE & 0.218 & 0.406 & 0.626 & 0.570 & 1.054 \\
GALAH & 0.059 & 0.102 & 0.163 & 0.119 & 0.255 \\
LAMOST & 0.104 & 0.154 & 0.236 & 0.197 & 0.332 \\
RAVE & 0.252 & 0.335 & 0.465 & 0.451 & 0.709 \\
\end{tabular}
\end{footnotesize}
\end{table}

\subsubsection{Surface gravity}

\begin{figure}
\begin{center}
\includegraphics[width=\columnwidth]{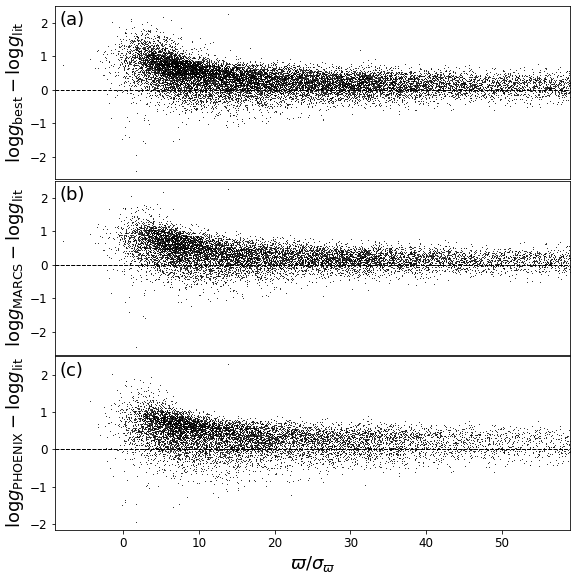}
\end{center}
\caption{Bias of surface gravity estimates for stars with low parallax signal-to-noise ratio, $\varpi/\sigma_\varpi$, for 25\,169 red clump stars \citep{ApogeeDR16,2014ApJ...790..127B}. Panel (a): \gspphot\ best-library results. Panel (b): \gspphot\ MARCS library results. Panel (c): \gspphot\ PHOENIX library results.}
\label{fig:red-clump-logg-gspphot-paper}
\end{figure}

Table~\ref{table:logg-comparison-to-literature} demonstrates that our $\log g$ estimates are also very good: as is evident from the median absolute differences, half of our sources agree with literature values of $\log g$ to within 0.25 dex. For GALAH DR3 in particular, 75\% of our results agree to within 0.12 dex. Given the low resolution of BP/RP, this may appear somewhat surprising, but the main constraint on $\log g$ is actually imposed by the prediction of the apparent $G$ magnitude together with the measured parallax that provide a rather tight constraint on the luminosity and absolute magnitude of the star. For sources with poor parallax measurements ($\frac{\varpi}{\sigma_\varpi}<10$), Fig.~\ref{fig:red-clump-logg-gspphot-paper} reveals that \gspphot\ tends to systematically overestimate $\log g$. This is related to an overly harsh distance prior, as is discussed in Sect.~\ref{ssect:distances_vs_prior}.

\begin{figure}
\begin{center}
\includegraphics[width=\columnwidth]{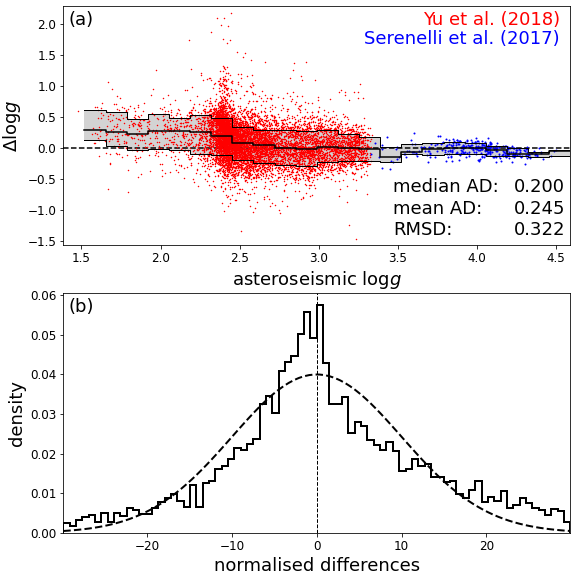}
\end{center}
\caption{Comparison to asteroseismic surface gravities. Panel (a): Comparison between our best-library $\log g$ estimates and asteroseismic values from \citet{Serenelli2017} (blue points) and \citet{Yu2018} (red points). The solid black line shows the median difference and the grey region the central 68\% interval. Panel (b): Distribution of normalised differences defined in Eq.~(\ref{eq:def:normalised-logg-residuals}) for asteroseismic $\log g>2.5$ compared to a Gaussian with zero mean and $\sigma=10$ (dashed black line).}
\label{fig:asteroseismic-logg-comparison}
\end{figure}

The excellent quality of our $\log g$ estimates for sources with good parallaxes is demonstrated further by the comparison to asteroseismic values in Fig.~\ref{fig:asteroseismic-logg-comparison}a. There is only a mild systematic overestimation by $\sim$0.25 dex for giants with $\log g<2.5$ dex. In particular, the median absolute difference quoted in Fig.~\ref{fig:asteroseismic-logg-comparison}a shows that half of our $\log g$ values agree with asteroseismic values to within 0.2 dex or better. 

Given that asteroseismic estimates of $\log g$ typically have very low uncertainties \citep{2013MNRAS.431.2419C}, the deviations in Fig.~\ref{fig:asteroseismic-logg-comparison} should be fully explained by \gspphot's own uncertainties. If we therefore normalise the differences by our uncertainty estimates, we can look for a distribution similar to a unit Gaussian. More specifically, we normalise by our lower confidence interval if our $\log g$ estimate is above the asteroseismic value and we normalise by our upper confidence interval if it is below. If we denote the normalised differences as $d$ and the asteroseismic reference value as $\log g_\textrm{AS}$, we have
\begin{equation}\label{eq:def:normalised-logg-residuals}
d=\left\{
\begin{array}{ll}
\frac{\textrm{logg\_gspphot} - \log g_\textrm{AS}}{\textrm{logg\_gspphot\_upper}-\textrm{logg\_gspphot}} & \Leftrightarrow \textrm{logg\_gspphot}<\log g_\textrm{AS} \\
\frac{\textrm{logg\_gspphot} - \log g_\textrm{AS}}{\textrm{logg\_gspphot}-\textrm{logg\_gspphot\_lower}} & \Leftrightarrow \textrm{logg\_gspphot}>\log g_\textrm{AS}
\end{array}
\right.
.\end{equation}
We emphasise that the lower and upper confidence levels are 16th and 84th percentiles of the MCMC samples and the reported value is the median (see Sect.~\ref{ssec:values-uncertainties-mcmc}) such that

$\textrm{logg\_gspphot\_lower}\leq\textrm{logg\_gspphot}\leq\textrm{logg\_gspphot\_upper,}$

and $\textrm{logg\_gspphot\_upper}-\textrm{logg\_gspphot}\geq 0$ is the upper error while $\textrm{logg\_gspphot}-\textrm{logg\_gspphot\_lower}\geq 0$ is the lower error.
We exclude sources for which the asteroseismic reference value is below 2.5\,dex, because our $\log g$ estimates are biased  in that regime. This systematic error would compromise the validation of our uncertainty estimates that are meant to account for random errors only. Unfortunately, Fig.~\ref{fig:asteroseismic-logg-comparison}b shows that while the distribution of normalised differences is indeed centred on zero, it is very far from a unit Gaussian. Instead, Fig.~\ref{fig:asteroseismic-logg-comparison}b suggests that our uncertainties are underestimated by a factor of $\sim$10. This value of 10 has been estimated by a maximum-likelihood fit of the standard deviation given the normalised residuals. Indeed, the distribution appears to be not even Gaussian at all, exhibiting a sharper peak and heavier tails than expected from a Gaussian. We return to this issue in Sect.~\ref{ssec:randomly-split-sources}.

\begin{table}
\caption{Comparison of the best [M/H] estimate from \gspphot\ to literature values. Same as Table~\ref{table:teff-comparison-to-literature}.}
\label{table:metal-comparison-to-literature}
\centering
\begin{footnotesize}
\begin{tabular}{l|r|r|r|r|r}
catalogue & MedAD & MAD & RMSD & AD 75\% & AD 90\% \\
\hline
APOGEE & 0.210 & 0.303 & 0.450 & 0.384 & 0.644 \\
GALAH & 0.210 & 0.238 & 0.295 & 0.326 & 0.452 \\
LAMOST & 0.204 & 0.248 & 0.330 & 0.328 & 0.477 \\
RAVE & 0.198 & 0.254 & 0.350 & 0.344 & 0.524 \\
\end{tabular}
\end{footnotesize}
\end{table}

\subsubsection{Metallicity}

Table~\ref{table:metal-comparison-to-literature} compares the \gspphot\ metallicity estimates to literature values. Despite the low resolution of BP/RP spectra, half of the sources agree with literature values of [M/H] to within 0.21. However, we caution that the [M/H] provided by  \gspphot \ are systematically too low, which is not obvious from Table~\ref{table:metal-comparison-to-literature}. While we would expect a systematic underestimation of [M/H] for sources with overestimated extinction due to the degeneracy between these parameters, we also observe that the \gspphot\  [M/H] are too low when the extinctions are adequately estimated. Consequently, the \gspphot\ [M/H] values should only be used with due caution. However, we find that \gspphot\ [M/H] estimates can be empirically calibrated to the [Fe/H] scale  of LAMOST DR6\footnote{\url{http://dr6.lamost.org/v2/catalogue}} \citep{2012RAA....12..723Z,2012RAA....12..735D,2015RAA....15.1089L} \linksec{ssec:cu8par_apsis_gspphot_results}{see also Sect.~\onlinedocucu8section.3.3.6 in the online documentation for details}.

We briefly outline this empirical metallicity calibration procedure here: Our objective is to use a multivariate adaptive regression spline \citep[hereafter MARS,][]{MARS} in order to learn a mapping from \gspphot's biased [M/H] to some well-established metallicity estimates. We considered various literature catalogues as possible training samples and eventually opted for LAMOST DR6 because it provides a broad range of metallicity values but does not probe too deeply into high-extinction regions in the Galactic disk.\footnote{Our preferred solution would have been to train an [M/H] calibration based on \gspspec\ results \citep{DR3-DPACP-186}. Unfortunately, due to \gspphot\ filtering and \gspspec\ flagging, the overlap of both \apsis\ modules has an insufficient number of low-metallicity stars.} As LAMOST provides [Fe/H] estimates, our MARS model not only needs to remove the systematic errors from the  \gspphot\ [M/H] but also to translate from [M/H] to [Fe/H]. As the metallicity bias in \gspphot\ also depends on stellar parameters, the input features of the MARS model include the effective temperature, surface gravity, the biased [M/H] value itself, and the extinction and reddening. It also includes Galactic latitude, which helps with the translation from [M/H] to [Fe/H].\footnote{The conversion from [M/H] to [Fe/H] strictly requires knowledge of [$\alpha$/Fe]. While [$\alpha$/Fe] is not available from \gspphot, [$\alpha$/Fe] varies between the Galactic plane and high latitudes such that the MARS model can infer an approximate conversion from the Galactic latitude.} The trained MARS model then provides the calibrated [Fe/H].

\begin{figure}
\begin{center}
\includegraphics[width=\columnwidth]{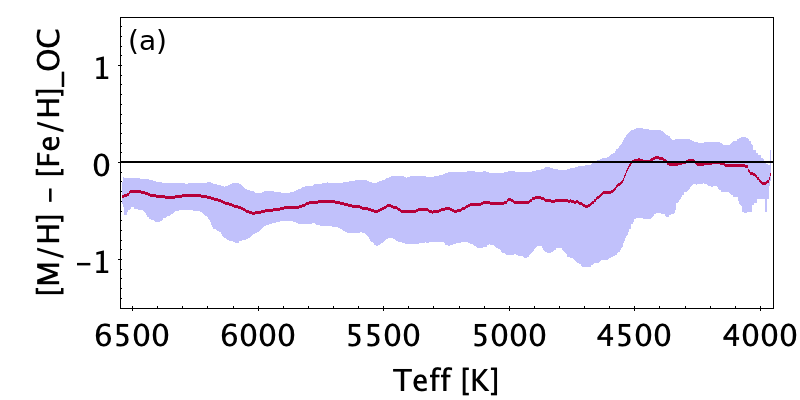}
\includegraphics[width=\columnwidth]{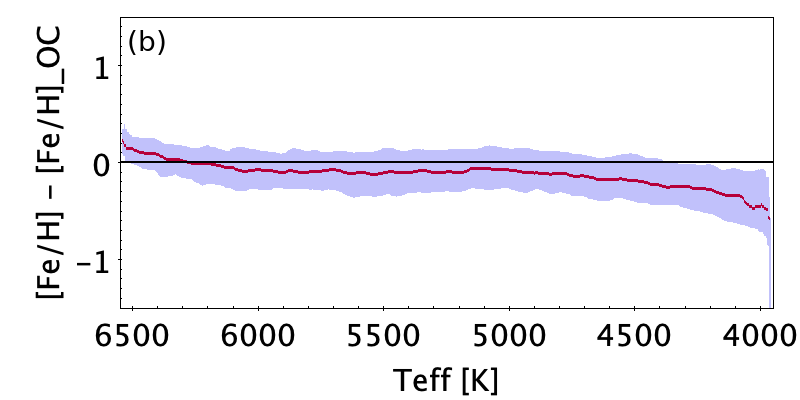}
\includegraphics[width=\columnwidth]{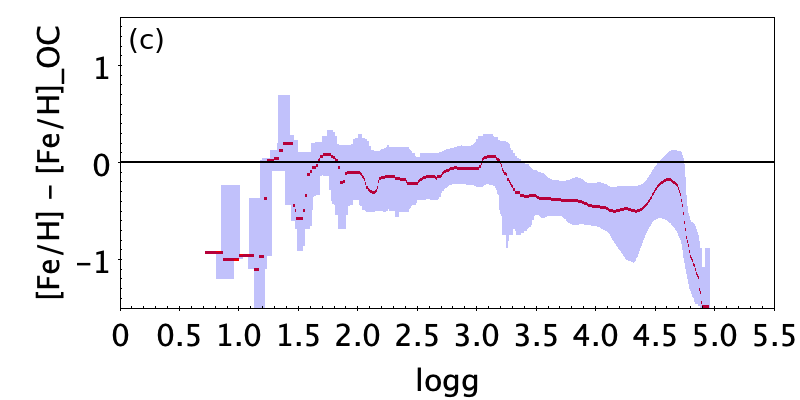}
\end{center}
\caption{Difference between \gspphot\ metallicities of individual FGK members and the mean [Fe/H] of the parent open cluster for stars with $\frac{\varpi}{\sigma_\varpi}\ge 10$ and MARCS models, before and after the calibration. We note that the calibration on LAMOST [Fe/H] values translates [M/H] to [Fe/H]. The red line is the median of the distribution. The blue area is delimited by the 16th and 84th quantiles. Panel (a): Residuals versus $T_\textrm{eff}$ before the calibration. Panel (b): Residuals versus $T_\textrm{eff}$ after the calibration. Panel (c): Residuals versus $\log g$ after the calibration.}
\label{fig:mh_calibration_open_cluster}
\end{figure}

We test this calibration with FGK members ($T_\textrm{eff}$ between 4000 and 6500~K) of open clusters with known metallicities after rejection of stars with poor parallax measurements ($\frac{\varpi}{\sigma_\varpi}<10$). Cluster members and mean cluster metallicities were taken from \citet{2021arXiv211207545S}. Figure~\ref{fig:mh_calibration_open_cluster}a shows the individual metallicities [M/H] minus reference [Fe/H] of the parent cluster as a function of the temperature of the  star. This test involves nearly 56\,000 stars in 187 open clusters. Figure~\ref{fig:mh_calibration_open_cluster}b shows that there is a net improvement of the metallicities when the calibration is applied ---in both the offset and the dispersion--- over the temperature range. However, the systematic errors are not completely removed by the calibration and still depend on the surface gravity (despite the MARS calibration model taking $\log g$ as an input feature), as is evident from Fig.~\ref{fig:mh_calibration_open_cluster}c.
Therefore, we emphasise that our empirical calibration is simply an illustration. The users are explicitly encouraged to find better calibration procedures of their own.
Nevertheless, as mentioned in \citet{DR3-DPACP-123}, the [M/H] provided by \gspphot\,, together with its estimates of temperature and gravity, can still be used to select solar-like stars whose RVS spectra are in close agreement with those of known solar analogues. Therefore, \gspphot\ [M/H] estimates, in spite of their large systematic errors, still contain some exploitable information about the actual metallicity of the star, which is why they were not removed from the \gdr3\ release.

This issue of [M/H] discrepancies is likely due to the mismatch between observed BP/RP spectra and the models employed by \gspphot\ (see Sect.~\ref{ssec:XP-model-mismatch}). Given the large differences between observed and model BP/RP spectra shown in Fig.~\ref{fig:solar-twins-model-mismatch}c and d, it is not surprising that the metallicity estimates are of poor quality. Metallicity is the weakest parameter in \gspphot, in the sense that it has the lowest impact on the shape of BP/RP spectra and so is most affected by model--data mismatches. In particular, the metallicity information is largely encoded at the blue end of the BP spectrum, where Fig.~\ref{fig:solar-twins-model-mismatch}c shows large discrepancies between observations and models.

\subsection{Extinction}

\gspphot\ extinction estimates are validated in various places. For example, \citet{DR3-DPACP-123} select solar-like stars from \gspspec\ results \citep[based on RVS spectra,][]{DR3-DPACP-186} and show that the $G_\textrm{BP}-W_2$ colour of those solar-like stars is in close  agreement with the linear trend with the \gspphot\ \abp\ estimate to within 0.087~mag RMS scatter.
\citet{DR3-DPACP-144} find good agreement between \gspphot's \ebpminrp\ reddening and the equivalent widths of diffuse interstellar bands (DIBs) measured from Gaia RVS spectra.
\gspphot\ extinction estimates are also used to estimate a map of total Galactic extinction and \citet{DR3-DPACP-158} report that this map agrees very well with Planck data for $A_0<4$~mag and is also in good agreement with the Schlegel map \citep{1998ApJ...500..525S}. In this section, we complement the aforementioned findings with some additional validation results.

\subsubsection{Local Bubble and non-negativity}

\begin{figure}
\begin{center}
\includegraphics[width=\columnwidth]{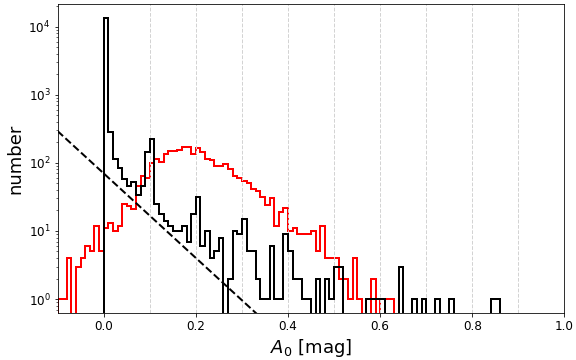}
\end{center}
\caption{Distribution of $A_0$ estimates for stars in the Local Bubble ($\varpi>20$ mas). The dashed black line indicates the slope of an exponential with 0.07 mag scale length. The vertical dashed grey lines indicate grid points of $A_0$ from multilinear interpolation of model spectra. (See Appendix~\ref{appendix:example-ADQL-queries} for the ADQL query.) The red histogram shows \texttt{av50} extinction estimates from StarHorse2021 \citep{2022A&A...658A..91A} with \texttt{sh\_outflag}=0000.}
\label{fig:local-bubble-extinction}
\end{figure}

\gspphot\ imposes the constraint $A_0\geq 0$, which reflects the fact that extinctions cannot be negative. This causes \gspphot\ to systematically overestimate $A_0$ in regimes where the actual extinction is very low \citep[also see TGE results in ][]{DR3-DPACP-158}. We illustrate this effect with the Local Bubble: \gdr3 contains 51\,983 sources with parallaxes larger than 20 mas (i.e.\ closer than 50 pc), which should have very low extinction. Of these, 14\,862 have \gspphot\ results\footnote{The high-quality parallaxes in the Local Bubble make it difficult for \gspphot\ to match the inverse distance within $10\sigma$ of the parallax measurement (see Sect.~\ref{ssec:filtering}), which means many sources are filtered out.} and the $A_0$ distribution is shown in Fig.~\ref{fig:local-bubble-extinction}. While the average extinction in this sample is $A_0=0.1$~mag, the values can be significantly larger. This is expected given that $A_0$ is subject to measurement noise. For a non-negative random variate with a true value of zero, we expect an exponential distribution given that this is the maximum-entropy distribution for such a random variate \citep[e.g.][]{DowsonWragg1973}. Indeed, Fig.~\ref{fig:local-bubble-extinction} shows that the $A_0$ distribution is roughly matched by an exponential with a scale length of 0.07~mag. Similarly to Fig.~\ref{fig:local-bubble-extinction}, we obtain exponentials with scale lengths of 0.07~mag for \abp, 0.06~mag for $A_G$, and 0.05~mag for \arp. These also provide rough error estimates, at least for bright sources in the Local Bubble. In particular, the 0.07~mag for \abp\ is consistent with the 0.087~mag RMS scatter in \abp\ reported for solar-like stars in \citet{DR3-DPACP-123}. However, we note that this is not purely random but includes some systematic errors.  Figure~\ref{fig:local-bubble-extinction} shows peaks at the grid points used for the multilinear interpolation of model spectra, in particular around $A_0$ values of 0, 0.1, and 0.2. As discussed in Sect.~\ref{ssec:XP-model-mismatch}, this is most likely a result of the mismatch between models and real BP/RP spectra. Furthermore, the extinction overestimation tends to affect some parts of the CMD more than others \citep[see][]{DR3-DPACP-127}. In particular, low-mass dwarfs tend to have their extinction overestimated, which may be related to our Hertzsprung-Russell diagram prior using isochrones that do not coincide with the PARSEC isochrones employed by \gspphot\ in this regime (see Sect.~\ref{sect:priors}).

\subsubsection{Comparison to StarHorse2021}

We briefly compare the \gspphot\ \ \azero\ estimate to the \texttt{av50} estimate from StarHorse2021 \citep{2022A&A...658A..91A}. Returning to the Local Bubble ($\varpi>20$ mas), the red histogram in Fig.~\ref{fig:local-bubble-extinction} shows that StarHorse2021 overestimates extinction with a mean \texttt{av50} of 0.16 mag, which is about twice as large as the value from \gspphot. In particular, despite StarHorse2021 allowing for slightly negative extinctions, the \texttt{av50} does not peak near zero.

\begin{figure}
\begin{center}
\includegraphics[width=\columnwidth]{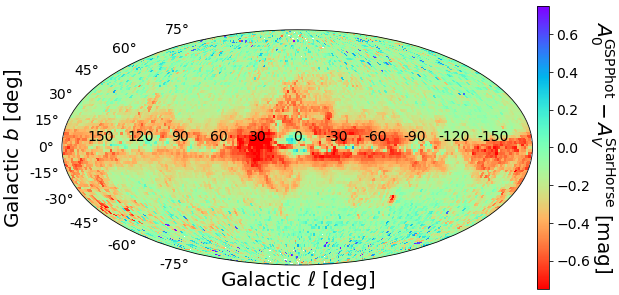}
\end{center}
\caption{Difference between the \gspphot\ $A_0$ and the \texttt{av50} provided by  StarHorse2021 with \texttt{sh\_outflag}=0000 \citep{2022A&A...658A..91A} on the sky. This skymap uses the Mollweide projection where lines of constant latitude are horizontal straight lines parallel to the equator.}
\label{fig:StarHorse2021-vs-GSPPhot-comparison-in-skymap}
\end{figure}

We also compare extinctions for a random subset of one million stars. On average,  \texttt{av50} estimated by StarHorse2021 is 0.36 mag higher than the \azero\ estimated by 
\gspphot. As is evident from Fig.~\ref{fig:StarHorse2021-vs-GSPPhot-comparison-in-skymap}, this difference is mainly driven by high-extinction regions. Outside such regions, StarHorse2021 \texttt{av50} appears to be about 0.1 mag higher than \gspphot\ \ \azero, which is very similar to the difference we find in the Local Bubble.  \citet{2022A&A...658A..91A} report in their Fig.~15 a systematic overestimation of \texttt{av50} in open clusters from \citet{2020A&A...640A...1C} that is consistent with the 0.1 mag difference to the \gspphot\  \azero.
However, in high-extinction regions, the differences can easily reach 0.7 mag or higher (Fig.~\ref{fig:StarHorse2021-vs-GSPPhot-comparison-in-skymap}). This systematic difference cannot be understood as \texttt{av50} being the Johnson $V$ band extinction and \azero\ being the monochromatic extinction at 541.4~nm. Even though the difference between these two extinction concepts becomes more pronounced as the extinction increases, the effect goes in the opposite direction, that is,\ \texttt{av50} should become increasingly smaller than \azero, not larger (\linksec{ssec:cu8par_inputdata_xp_SimuExtLaw}{see Sect.~\onlinedocucu8section.2.3.1.4 in the online documentation for details}).

\subsubsection{Comparison to Bayesstar19}

\begin{figure*}
\begin{center}
\includegraphics[width=0.33\textwidth]{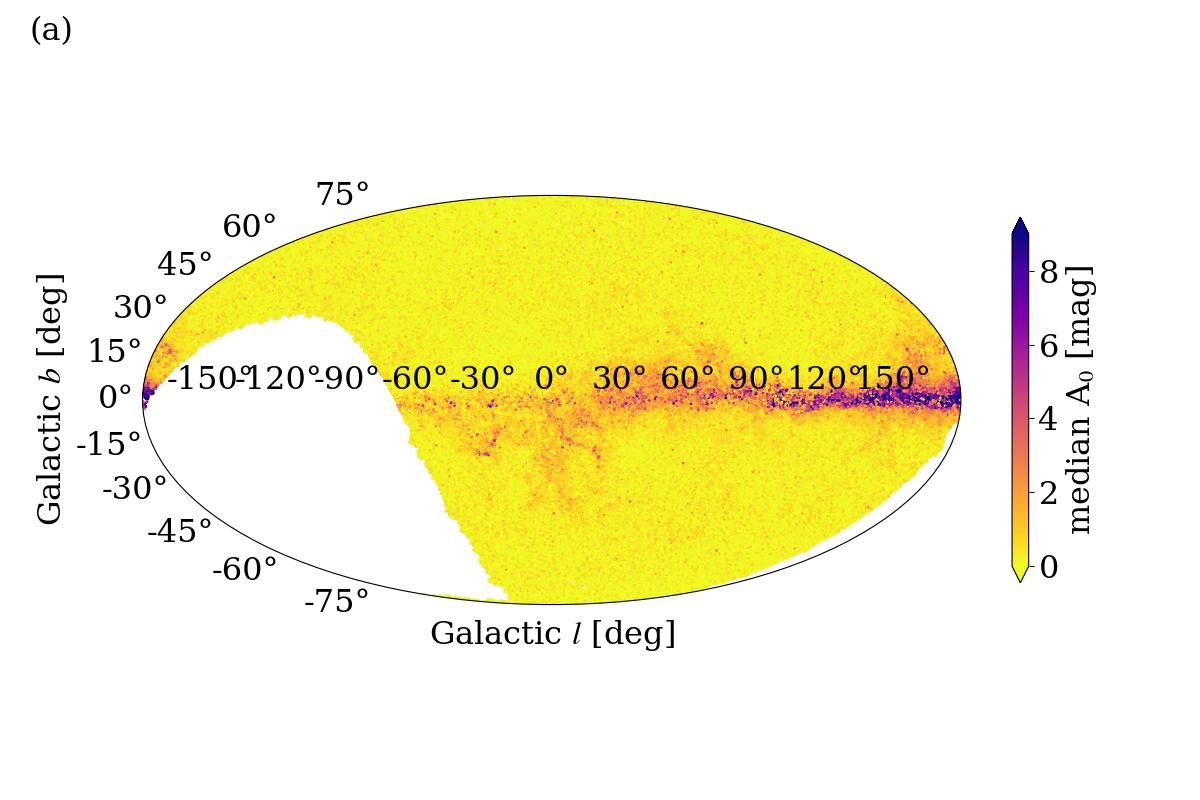}
\includegraphics[width=0.33\textwidth]{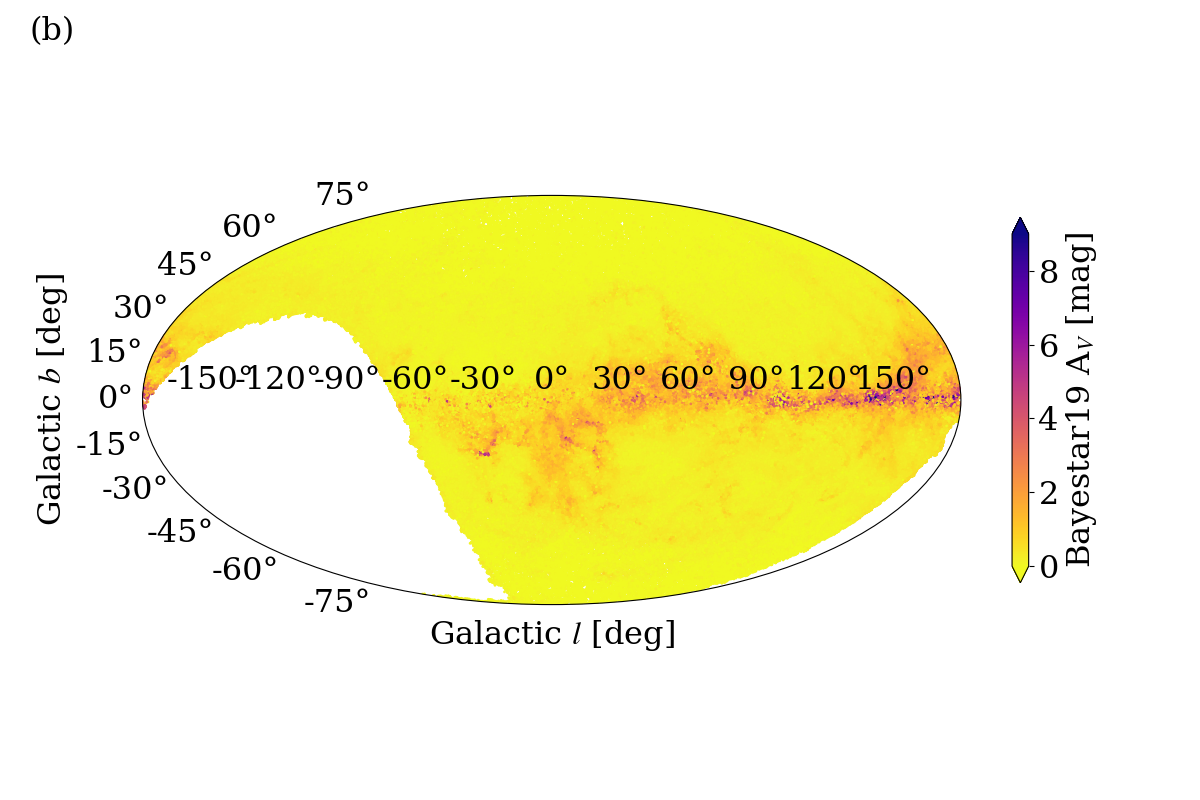}
\includegraphics[width=0.33\textwidth]{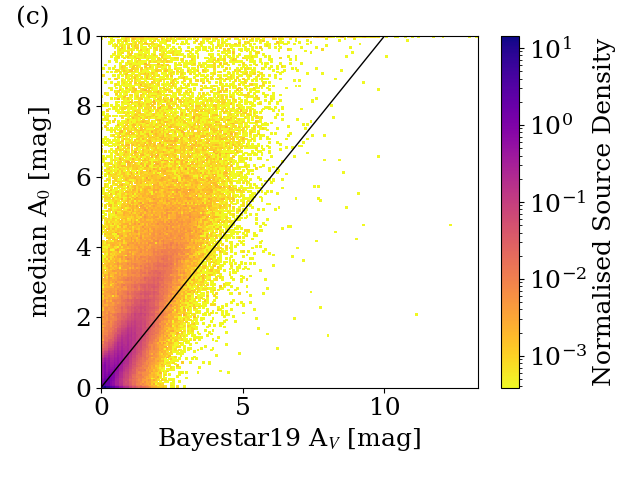}
\end{center}
\caption{Extinction comparison between \gspphot\ and Bayestar19 for a randomly selected sample of 1 million sources. Panel (a): Skymap of\  \azero\ provided by  \gspphot, \ taking a pixel-wise median value. Panel (b): Skymap of  $A_{V}$ provided by
Bayestar19, taking a pixel-wise median value. Panel (c): One-to-one comparison of the \gspphot\ median \azero\ and the\ Bayestar19 $A_{V}$. Both skymaps use the Mollweide projection where lines of constant latitude are horizontal straight lines parallel to the equator. All panels show the identical sample of stars. The missing data in panels (a) and (b) are due to the footprint of Bayestar19.}
\label{fig:Baye19ExtComp_Rand1mil}
\end{figure*}

\begin{figure*}
\begin{center}
\includegraphics[width=0.33\textwidth]{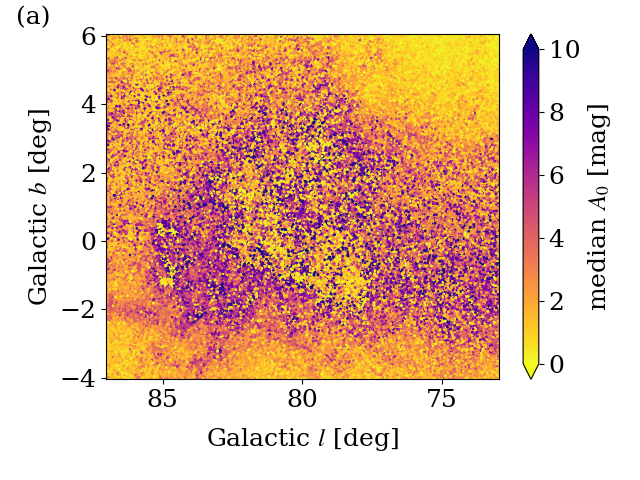}
\includegraphics[width=0.33\textwidth]{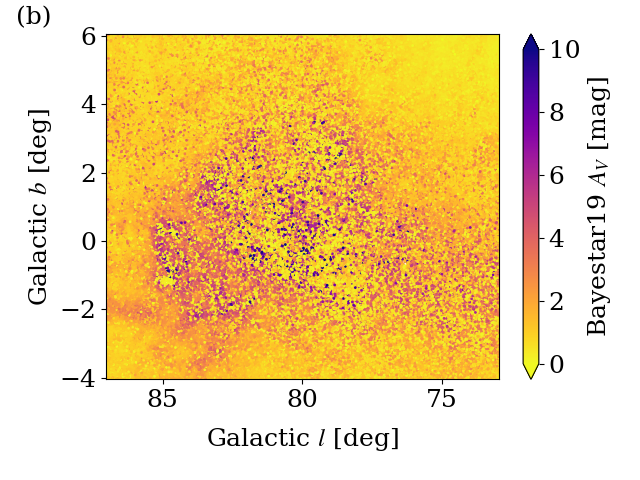}
\includegraphics[width=0.33\textwidth]{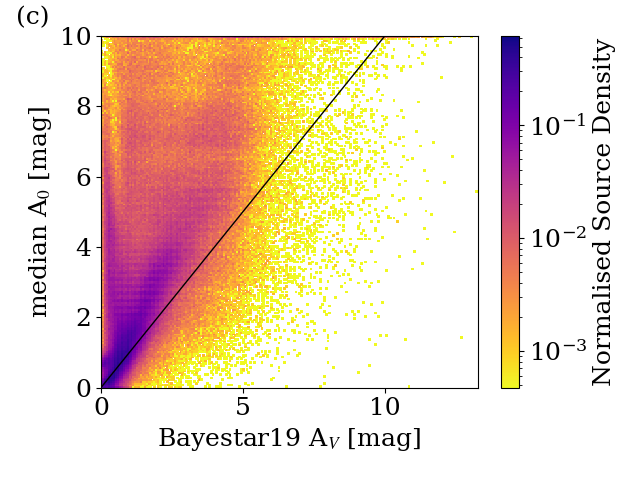}
\end{center}
\caption{Same as Fig.~\ref{fig:Baye19ExtComp_Rand1mil} but for the dense star-formation region Cygnus X1.}
\label{fig:ExtComp_CygX}
\end{figure*}

For further validation, we compare the \gspphot\ \ \azero\ extinctions to the $A_{V}$ extinctions derived from the Bayestar19 3D extinction map \citep{Green2019_Bayestars19}. As mentioned above, the \gspphot\ \azero\ is the monochromatic extinction at 541.4~nm and the parameter in our extinction law \citep{1999PASP..111...63F}. From Bayestar19, $A_{V}$ is conceptually the closest to \gspphot\  \azero. We sample the 3D extinction map from Bayestar19 using the \gspphot\ distance to each star. We define two subsamples for comparison: (1) a randomly selected sample of 1 million sources spread throughout the sky, and (2) all the sources in the direction of the Cygnus~X star-formation region, $73 \leq l \leq 87, -4 \leq b \leq 6$;
these samples are shown in Fig.~\ref{fig:Baye19ExtComp_Rand1mil} and Fig.~\ref{fig:ExtComp_CygX}.
In both samples, \gspphot\ predicts a higher extinction than Bayestar19, as is evident from the all-sky sample in Fig.~\ref{fig:Baye19ExtComp_Rand1mil}c as well as Cygnus~X in Fig.~\ref{fig:ExtComp_CygX}c. The differences between \gspphot\ and Bayestar19 extinctions become larger as the extinctions increase. However, as reported in \citet{DR3-DPACP-158}, Bayesstar19 appears to also estimate lower extinctions compared to data from Planck and Schlegel. We note that a comparison between Fig.~\ref{fig:Baye19ExtComp_Rand1mil}a and b shows that the \gspphot\ map shows finer structures with larger contrast (i.e.\ higher extinctions). We put forward a possible explanation for this below.

Looking at the Cygnus~X star-formation region, \gspphot\ appears to more faithfully recover extinctions towards compact high-density regions when compared to Bayestar19. This is evident from comparing Fig.~\ref{fig:ExtComp_CygX}a and b, where regions of significantly higher extinction are visible in \gspphot\ results, tracing the structure of the dense regions of ongoing star formation. In Fig.~\ref{fig:ExtComp_CygX}c, we directly compare the extinctions from \gspphot\ and Bayestar19. We see two clear populations of sources: the majority of sources have similar extinctions in the two catalogues, although \gspphot\ is systematically higher, while a smaller population have large extinctions in \gspphot\ but negligible extinction predicted by Bayestar19. The first of these populations can be tied to stars in diffuse regions. The second population, where \gspphot\ predicts large extinctions but Bayestar19 does not, is only seen in regions that have high ISM densities, meaning \gspphot\  successfully recovers stars in regions with large dust density and active star formation while Bayestar19 does not.\footnote{We double-checked that these stars driving the high extinction values are {not} low-mass dwarfs at the faint/cool end of the main sequence which can sometimes exhibit spuriously large extinctions, e.g.\ in the Local Bubble \citep{DR3-DPACP-127}. Instead, these high-extinction values in Cygnus X1 appear to be driven by red giant stars.}

We notice that \gspphot\ maps show finer structures and larger extinction values than maps from Bayesstar19 in Fig.~\ref{fig:Baye19ExtComp_Rand1mil}a and b and Fig.~\ref{fig:ExtComp_CygX}a and b. We speculate that while Bayestar19 is capable of detecting high-extinction sources, the Gaussian process model the authors applied essentially smoothes out the map and averages over many lines of sight in its grid pixel. Therefore, a high-extinction line of sight may become averaged down because it occupies a small volume and therefore only affects a small fraction of sources. \gspphot\ on the other hand is not biased by this and recovers an extinction for each source individually, leaving the high-extinction sources as high extinction. While small-scale structures also exist in low-extinction regions, the bias from averaging will be more visible in high-dust-density, active star-formation regions.  Those small-scale structures may be washed out by the Gaussian process model employed by Bayestar19.

As an aside, we mention that both Fig.~\ref{fig:Baye19ExtComp_Rand1mil}c and Fig.~\ref{fig:ExtComp_CygX}c show horizontal stripes in the \gspphot\ \azero\ estimates. These are the same linear interpolation issues that we already observed in the Local Bubble (Fig.~\ref{fig:local-bubble-extinction}) and which are most likely caused by the mismatch between models and real BP/RP spectra (see Sect.~\ref{ssec:XP-model-mismatch}).

\subsection{Systematic underestimation of distances}
\label{ssect:distances_vs_prior}

\citet{DR3-DPACP-160} show that \gspphot\ distances of cluster member stars are consistent with the cluster distances from \citet{2020A&A...640A...1C} only out to 2-3~kpc, and that \gspphot\ distances become systematically too low beyond 3~kpc. Similar results are reported for stars with asteroseismic distances by \citet{Huber2017} and \citet{Anders2017}. However,  \citet{DR3-DPACP-160} also show that \gspphot\ distances are reliable out to 10~kpc for stars with high-quality parallax measurements ($\varpi/\sigma_\varpi\geq 10$). This dependence on parallax quality suggests that the systematic underestimation of distances may be related to the distance prior.
As we note in Sect.~\ref{sect:priors}, the length scale of the distance prior  is set to one-tenth of the length scale that we compute from the \edr3\ mock catalogue of \citet{Rybizki2020}. The objective to do so was to reduce the differences to literature values, for example, for effective temperatures. Unfortunately, this distance prior is overly harsh, resulting in a systematic underestimation of distances by \gspphot\ for sources with low parallax quality. This may also compromise other parameters too, such as the $\log g$ estimates of red clump stars in Fig.~\ref{fig:red-clump-logg-gspphot-paper}. This view is also supported by Fig.~\ref{fig:APOGEE-Teff-best-residuals-in-HRD}, which demonstrates that restricting to high parallax quality stabilises \gspphot\ and, in that case, also improves the temperature estimates.

\begin{figure}
\begin{center}
\includegraphics[width=\columnwidth]{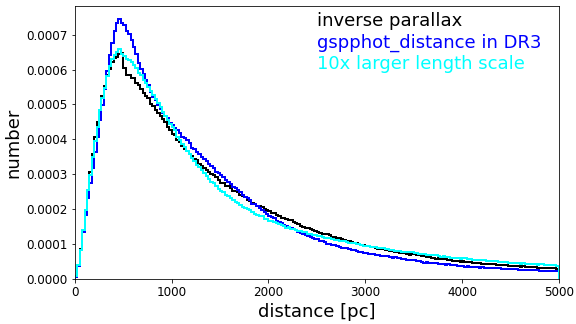}
\end{center}
\caption{Distributions of inverse parallax (black histogram), \gspphot\ distances in Gaia DR3 with overly harsh distance prior (blue histogram), and \gspphot\ distances after reprocessing with a relaxed distance prior (cyan histogram). The parallaxes shown here and also used during the \gspphot\ processing all include the zero-point correction from \citet{2021A&A...649A...4L}.} 
\label{fig:distance-distributions-priors}
\end{figure}

In order to confirm this interpretation, we locally reprocess 5 million sources with the length scale of the distance prior  relaxed by a factor of ten, that is,\ restoring the value we compute from the \edr3\ mock catalogue of \citet{Rybizki2020}. Unfortunately, these 5 million sources are not representative of the sample as a whole, but other sources are not available for this exercise for various reasons. Figure~\ref{fig:distance-distributions-priors} shows that while the \gspphot\ distances in \gdr3\ do not follow the inverse parallax distribution very well, the situation clearly improves when we relax the distance prior. We caution though that inverting parallaxes is not recommended \citep[e.g.][]{2015PASP..127..994B}, in particular given that many sources in this sample may have very noisy parallax measurements.
Nevertheless, given this systematic underestimation of distances in \gdr3\ for stars with low parallax quality, \gspphot\ distances cannot be used to map the Milky Way spiral arms \citep{DR3-DPACP-75}, the spatial distributions of the diffuse-interstellar-band absorption \citep{DR3-DPACP-144}, or chemical cartography \citep{DR3-DPACP-104}.

\subsection{Further validation results}

As mentioned above, \gspphot\ results are validated in various publications accompanying \gdr3. Here, we want to briefly highlight some of these findings:
In \citet{DR3-DPACP-160}, we show that using the radius and distance from \gspphot, we can predict angular diameters that are in excellent agreement with measurements from ground-based interferometry.
In \citet{DR3-DPACP-157}, we show that results from \gspphot\ and \flame\ are in very good agreement for radii, luminosity, and bolometric correction. We also report relatively good agreement in terms of effective temperatures and extinctions between \gspphot\ and \esphs\ for hot stars ($T_\textrm{eff}>7500$K), which is used for the OB sample definitions in \citet{DR3-DPACP-75} and \citet{DR3-DPACP-123}.
\citet{DR3-DPACP-186} report good agreement between results from \gspphot\ (low-resolution BP/RP spectra) and \gspspec\ (RVS spectra) for effective temperatures and surface gravities.
\citet{DR3-DPACP-123} demonstrate that when selecting solar-like stars from \gspspec\ results, the  colours of the resulting candidates are in good agreement with those of known solar twins for stars where $A_0<0.001$mag according to \gspphot. Furthermore, BP/RP spectra of solar-like stars exhibit a clear dimming and reddening trend with increasing $A_0$.
\citet{DR3-DPACP-79} find that temperature uncertainties from \gspphot\ are too small by a factor of approximately 4 for $\delta$Scuti and $\gamma$Doradus stars (spectral type early-F to mid-A), as well as for hotter variable stars, such as SPB or $\beta$Cephei (spectral type B9 or hotter).
Finally, \citet{DR3-DPACP-127} provide an overview of the main issues identified in \gdr3\ data, including \gspphot\ results.

\subsection{Uncertainty validation}
\label{ssec:randomly-split-sources}

It is conceptually very challenging to validate uncertainty estimates because we not only need reliable reference values but also reliable uncertainties on these reference values. In Fig.~\ref{fig:asteroseismic-logg-comparison} we can validate our $\log g$ uncertainties assuming that measurement errors in asteroseismic gravities are negligible compared to \gspphot\ uncertainties \citep[e.g.][]{2013MNRAS.431.2419C}.

We further validate our uncertainty estimates by employing the BP/RP split-epoch validation dataset introduced in \citet{EDR3-DPACP-118}: first, we go back to the epoch BP/RP spectra of each source\footnote{In \gdr3, each source typically has $\sim$40 epoch BP/RP spectra, but this can vary between 10 and over 150.} and randomly group them into two sets; second, for both sets of epoch spectra we compute a mean BP/RP spectrum. This procedure provides two statistically independent BP/RP spectra for each source. Each BP/RP spectrum now only has half of the epochs of the actual source, and so~this procedure produces spectra with slightly lower signal-to-noise ratio. As both spectra belong to the same source, the parameters obtained from processing both BP/RP spectra with \gspphot\ must be consistent with each other within their respective uncertainties. This should even be the case for intrinsically variable sources because the splitting of epoch spectra is done randomly. We note that we do not need to know the true parameters of each source; it is sufficient to know that the two randomly split spectra belong to the same source. As it is too time-consuming to perform this test for all sources, we can only do it for a small sample of 17 994 sources for which the necessary epoch BP/RP spectra were still available. This sample is not representative but still covers the apparent $G$ magnitude range reasonably well. We also require that \gspphot\ results for both components pass the filters described in Sect.~\ref{ssec:filtering}.

\begin{figure*}[h!]
\begin{center}
\includegraphics[width=2\columnwidth]{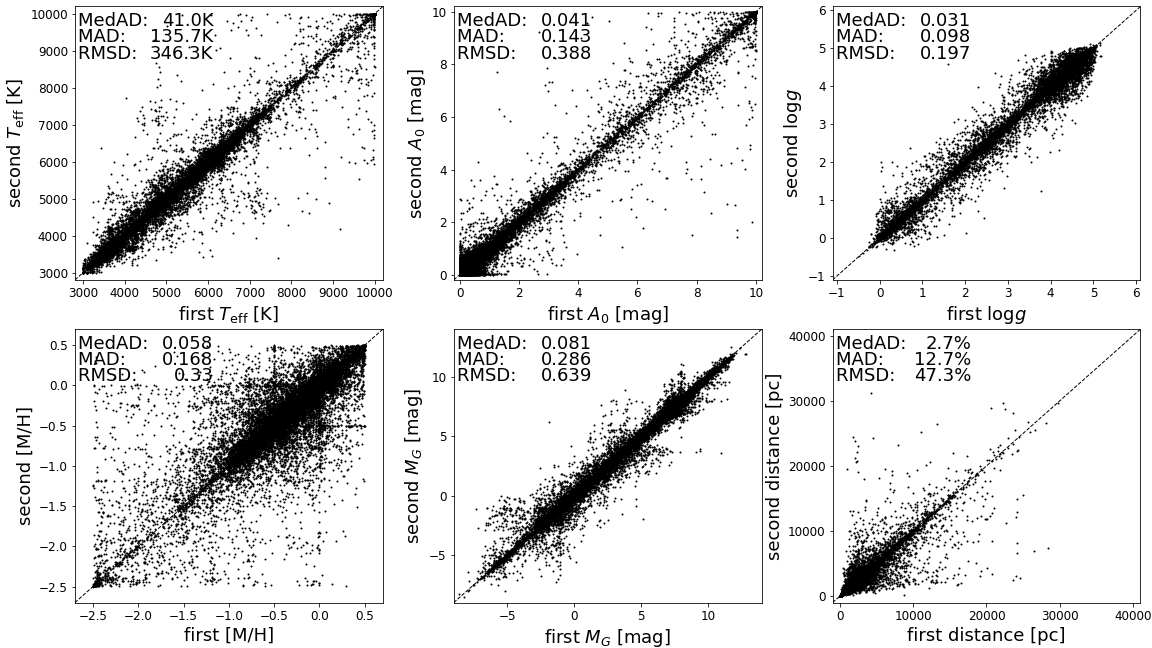}
\end{center}
\caption{Comparison of parameters between two components of 17\,994 sources from the BP/RP split-epoch validation dataset. Quoted numbers summarise the median absolute difference (MedAD), the mean absolute difference (MAD), and the root-mean-square difference (RMSD).}
\label{fig:AP-comparison-split-sources-DR3}
\end{figure*}

As a first test, Fig.~\ref{fig:AP-comparison-split-sources-DR3} simply compares the estimates within each pair. The median absolute differences are 41 K for effective temperature, 0.041 mag for extinction $A_0$, 0.031 for gravity, 0.058 for metallicity, 0.081 mag for absolute $M_G$ magnitude, and 2.7\% for distance. These differences are purely due to random noise and are much smaller than the differences with respect to literature values (e.g.~the median absolute differences between temperatures and literature values range from 110 to 170 K in Table~\ref{table:teff-comparison-to-literature}). This suggests that the differences from literature values are not driven by random errors but are rather dominated by systematic errors such as different temperature scales in \gspphot\ and other surveys.

If we compare the differences between the parameters to their uncertainty estimates, we find that, for 30\%-60\% of randomly split pairs, the parameters of the two components are outside each other's 68\% confidence intervals. In reality, this should only happen in 16\% of cases, which suggests that our uncertainties are systematically underestimated.

\begin{figure}[h!]
\begin{center}
\includegraphics[width=\columnwidth]{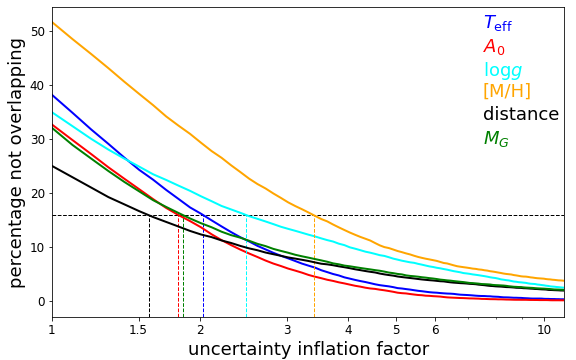}
\end{center}
\caption{Percentage of parameter pairs (in BP/RP split-epoch validation dataset) outside each other's 68\% confidence interval as a function of uncertainty inflation factor. The horizontal black dashed line at 16\% shows the expected percentage if uncertainties were correctly estimated. Vertical colour dashed lines indicate the inflation factors necessary for each parameter in order to bring the percentage of non-overlapping intervals to the expected 16\% level.}
\label{fig:DIDREQ-442-uncertainty-inflation-factors}
\end{figure}

In order to assess by how much our uncertainties are underestimated, Fig.~\ref{fig:DIDREQ-442-uncertainty-inflation-factors} investigates how the fraction of non-overlapping uncertainty intervals from \gspphot\ between the pairs decreases as we inflate the uncertainty intervals. The inflation is done by applying the same factor to the asymmetric intervals on both sides of each parameter. Table~\ref{table:uncertainty-inflation-factor} suggests that the uncertainties are systematically too small by factors ranging from 1.6 for distance to 3.4 for metallicity. However, we have to caution that the 17\,994 sources from the BP/RP split-epoch validation dataset may not be large enough or representative enough, meaning that the values in Table~\ref{table:uncertainty-inflation-factor} can only provide a rough indication. In particular, for asteroseismic gravities, Fig.~\ref{fig:asteroseismic-logg-comparison}  suggests that our uncertainties are too small by a factor of 10. Given this somewhat unclear situation, we do not apply any correction to the \gspphot\ uncertainties in the published \gdr3\ data.

\begin{table}[]
\centering
\caption{Inflation factors necessary to make the uncertainty intervals of randomly split pairs overlap in about 84\% of cases inferred from Fig.~\ref{fig:DIDREQ-442-uncertainty-inflation-factors}.}
\label{table:uncertainty-inflation-factor}
\begin{tabular}{c|c|c|c|c|c|c}
parameter & $T_\textrm{eff}$ & $A_0$ & $\log g$ & [M/H] & distance & $M_G$ \\
\hline
factor & 2.0 & 1.8 & 2.5 & 3.4 & 1.6 & 1.9
\end{tabular}
\end{table}

\subsection{Temperature--extinction degeneracy}
\label{ssec:teff-a0-degeneracy-randomly-split-sources}

\begin{figure}[h!]
\begin{center}
\includegraphics[width=\columnwidth]{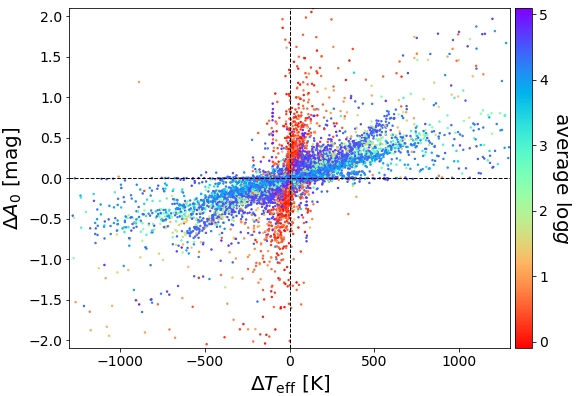}
\end{center}
\caption{Illustration of the temperature--extinction degeneracy using the parameter differences between pairs from the BP/RP split-epoch validation dataset (c.f.~Sect.~\ref{ssec:randomly-split-sources}). Colour-coding is done by the average surface gravity of both components.}
\label{fig:DIDREQ-442-Teff-A0-degeneracy}
\end{figure}

As a final use case of the BP/RP split-epoch validation dataset, Fig.~\ref{fig:DIDREQ-442-Teff-A0-degeneracy} clearly illustrates the degeneracy between effective temperature and line-of-sight extinction by comparing the differences in parameters for each pair. This degeneracy originates from the fact that low-resolution, optical BP/RP spectra are very similarly affected by both parameters. Figure~\ref{fig:DIDREQ-442-Teff-A0-degeneracy} also shows that the temperature--extinction degeneracy affects dwarfs and giants in different ways: while dwarfs can exhibit temperature variations as large as several hundred Kelvin even for small extinction variations, giants usually exhibit smaller temperature variations that are accompanied with much larger extinction variations. We note in particular that the temperature--extinction degeneracy works both ways in Fig.~\ref{fig:DIDREQ-442-Teff-A0-degeneracy}, causing simultaneous underestimation of $T_\textrm{eff}$ and $A_0$ just as frequently as a simultaneous overestimation.

\section{Summary}
\label{sect:summary}

In Gaia DR3, one of the major new data products is a collection of 220 million low-resolution BP/RP spectra. In this paper, we explain how the CU8 \apsis\ module \gspphot\ provides a homogeneous catalogue of stellar-parameter estimates for \numberofresults~million sources with $G<19$ based on these BP/RP spectra, parallax, and integrated photometry. We emphasise that \gspphot\ assumes that each source is a single star and that using combined BP/RP spectra implies that any intrinsic time variability is lost \citep{EDR3-DPACP-118}. One of the main design features of \gspphot\ is to {not} normalise the BP/RP spectra but instead to exploit the apparent flux level of the BP/RP spectra as an observational constraint on the radius and distance of the star (see Eq.~(\ref{eq:XP-amplitude})). \gspphot\ also employs PARSEC isochrones in order to guarantee astrophysically self-consistent stellar temperatures, gravities, metallicities, radii, and absolute magnitudes (see Sect.~\ref{sect:forward-isochrones}).

However, in \gdr3, \gspphot\ does {not} directly account for parallax and apparent $G$ magnitude when solving for the distance from the amplitude of BP/RP spectra (parallax and apparent magnitude still enter the \gspphot\ likelihood function in Eq.~(\ref{eq:gspphot-likelihood-function})). While this will be fixed in the future, in \gdr3 this requires filtering out \gspphot\ results where the distance is inconsistent with the measured parallax or the apparent $G$ magnitude (see Sect.~\ref{ssec:filtering}). In particular, this filtering removes \gspphot\ results for virtually all white dwarfs. Furthermore, \gspphot\ results have largely been filtered out for sources with negative parallaxes. Otherwise, this filtering does not appear to specifically affect any particular stellar population (see Fig.~\ref{fig:CMD-impact-filtering}) and \gspphot\ results are usually complete at the 80\% level except for sources brighter than $G<13$ or sources with extremely high parallax qualities (see Fig.~\ref{fig:filtering-completeness-vs-Gmag}). 

Despite the filtering, the \gspphot\ results remaining in \gdr3 can still suffer from several systematic effects \citep[an overview of the main issues identified in the \gdr3\ data is given by ][]{DR3-DPACP-127}:
\begin{enumerate}
\item \gspphot\ systematically underestimates distances for sources with low parallax quality, which applies to most sources in \gdr3. As discussed in Sect.~\ref{ssect:distances_vs_prior}, this is due to an overly harsh distance prior employed by \gspphot. Sources with high-quality parallax measurements ($\varpi/\sigma_\varpi>20$) should have reliable distances.
\item \gspphot\ imposes a non-negativity constraint on extinction. As a result, in low-extinction regions, \gspphot\ tends to overestimate extinction (e.g.\ in the Local Bubble as shown in Fig.~\ref{fig:local-bubble-extinction}).
\item The [M/H] estimates from \gspphot\ are dominated by large systematic errors which reduce them to the level of qualitative information. We therefore advise against using them. However, the  [M/H] estimates provided by \gspphot\ are still sufficiently informative that they can be empirically calibrated onto the LAMOST DR6 [Fe/H] scale, as we illustrate in Fig.~\ref{fig:mh_calibration_open_cluster}.
\item Given that BP/RP spectra have very low resolution and only cover the wavelength range from 320 to 1050 nm, there is a degeneracy between effective temperature and line-of-sight extinction: Increasing the star's effective temperature can be compensated by simultaneously increasing the line-of-sight extinction, thereby producing very similar BP/RP spectra. The strength of this temperature--extinction degeneracy varies with stellar population: while main sequence dwarfs can exhibit temperature variations as large as several hundred Kelvin even for small extinction variations, red giant stars usually exhibit smaller temperature variations that
are accompanied with much larger extinction variations (see  Fig.~\ref{fig:DIDREQ-442-Teff-A0-degeneracy}). This is a major limitation for \gspphot. We also point out that the temperature--extinction degeneracy, in principle, works both ways, but in low-extinction regimes tends to primarily cause an overestimation of temperatures and extinctions simply because the non-negativity constraint leaves no room to underestimate extinction.
\end{enumerate}
Another fundamental limitation of \gspphot\ in \gdr3 is a mismatch between observed BP/RP spectra and models. We illustrate this using solar-like stars in Fig.~\ref{fig:solar-twins-model-mismatch}. This mismatch is likely responsible for the poor quality of \gspphot\ \ [M/H] estimates, given that metallicity has the weakest impact on the shape of BP/RP spectra and is therefore easiest to compromise. At the aesthetic level, this mismatch also causes stripes in \gspphot\ results (e.g.\ Fig.~\ref{fig:ExtComp_CygX}c). As we discuss in Sect.~\ref{ssec:XP-model-mismatch}, this mismatch is unlikely to originate from the CU5 instrument model, but rather different solar model SEDs result in BP/RP spectra whose differences are easily measurable from \gdr3 data. While this is unfortunate for \gspphot, it will allow the community to further refine stellar atmospheric models.

Given all the aforementioned limitations, \gspphot\ results still compare well to expected values: In a comparison with GALAH DR3 and LAMOST DR4, half of the stars have temperatures that deviate by less than 110 K from the literature values (see Table~\ref{table:teff-comparison-to-literature}). The differences are larger for APOGEE DR16, which probes deeper into distant stars in the high-extinction regimes of the Galactic disk. If we restrict the comparison to APOGEE values to high-quality parallaxes ($\varpi/\sigma_\varpi>20$), we obtain results that are just as good as for GALAH DR3 or LAMOST DR4. Concerning surface gravities, half of the stars deviate by less than 0.25 from literature values (see Table~\ref{table:logg-comparison-to-literature}). A comparison to asteroseismic gravities confirms a median absolute difference of 0.2 (see Fig.~\ref{fig:asteroseismic-logg-comparison}a). Concerning extinctions, the Local Bubble suggests typical uncertainties of 0.07 mag in \azero and \abp, and slightly lower uncertainties of 0.06 mag in $A_G$ and 0.05 mag in \arp, reflecting the different susceptibilities to extinction of each band. This also agrees with the scatter of 0.087 mag in \abp\ that we find in solar-like stars in \citet{DR3-DPACP-123}.
A comparison to StarHorse2021 shows that, firstly, there is a global offset of \texttt{av50} that is about 0.1~mag larger than the  \gspphot\   \azero(e.g.\ at high latitudes and in the Local Bubble). This offset is likely due to a systematic overestimation of \texttt{av50}, which is also evident from the comparison to open clusters made by \citet{2022A&A...658A..91A}. Secondly, in high-extinction regions, the StarHorse2021 \texttt{av50} can be substantially larger than the \gspphot \ \azero  (see Fig.~\ref{fig:StarHorse2021-vs-GSPPhot-comparison-in-skymap}). This effect cannot be explained by the different definitions of \texttt{av50} and \azero\ (which should work in the exact opposite direction). The systematic differences between \gspphot\ and StarHorse2021 extinction estimates are currently not understood.

We caution that the uncertainty estimates from \gspphot\ tend to be much smaller than the typical differences from reference values. Validation of uncertainties is very difficult. In particular, a simple comparison to literature values is insufficient because the literature values also often have underestimated uncertainties. In this work, we circumvent this problem by producing two statistically independent incarnations of a limited sample of stars (the BP/RP split-epoch validation dataset discussed in Sect.~\ref{ssec:randomly-split-sources}). For these, we do not need to know their true stellar parameters. Instead, it is sufficient to know that both incarnations represent the exact same star, such that the \gspphot\ results for both should be consistent within their respective uncertainties. Unfortunately, we find that this is not the case, that is,\ \gspphot\ uncertainties are systematically underestimated by factors ranging from 1.6 to 3.4 (see Table~\ref{table:uncertainty-inflation-factor}). Likewise, uncertainties on surface gravities appear to be underestimated by a factor of $\sim$10, as is evident from a comparison to asteroseismic values (see Fig.~\ref{fig:asteroseismic-logg-comparison}b). There are multiple reasons why \gspphot\ underestimates uncertainties: Firstly, as mentioned above, some priors are too harsh and thereby may overly restrict the fit procedure. Secondly, the CU8 \apsis\ chain ignores correlations between pixels even though they exist \citep[e.g.][]{DR3-DPACP-157}. Finally, the aforementioned mismatch between observed BP/RP spectra and models not only causes systematic errors, but when the fit struggles to make the models match the observed data, it usually also leads to an underestimation of uncertainties.

\begin{acknowledgements}
This work presents results from the European Space Agency (ESA) space mission \gaia. \gaia\ data are being processed by the \gaia\ Data Processing and Analysis Consortium (DPAC). Funding for the DPAC is provided by national institutions, in particular the institutions participating in the \gaia\ MultiLateral Agreement (MLA). The \gaia\ mission website is \url{https://www.cosmos.esa.int/gaia}. The \gaia\ archive website is \url{https://archives.esac.esa.int/gaia}.
Acknowledgements are given in Appendix~\ref{ssec:appendixAcknowledgements}
\end{acknowledgements}

% WARNING
%-------------------------------------------------------------------
% Please note that we have included the references to the file aa.dem in
% order to compile it, but we ask you to:
%
% - use BibTeX with the regular commands:
%   \bibliographystyle{aa} % style aa.bst
%   \bibliography{Yourfile} % your references Yourfile.bib
%
% - join the .bib files when you upload your source files
%-------------------------------------------------------------------
\bibliographystyle{aa} % style aa.bst
\bibliography{bibliography,dr3papers,gaia_papers} % your references

\appendix

\section{Technical details}

\subsection{Deriving the prior distributions}
\label{appendix:priors-derivation}

In this Appendix, we derive the factorised priors step by step. Again, we emphasise that while priors are usually defined for actual fit parameters, they can just as well be defined for derived parameters instead. This may be uncommon but we find it easier to impose a prior in the Hertzsprung-Russell diagram (temperature vs. absolute magnitude, both derived parameters) than over the fit parameters of initial mass and age. For a given BP/RP spectrum $\vec s$, apparent $G$ magnitude, and parallax $\varpi$, the following posterior distribution is provided in \gdr3:
\begin{displaymath}
    P(T_\textrm{eff},\log g,[M/H],A_0,A_G,d,R,M_G|\vec s,G,\varpi)
\end{displaymath}
\begin{displaymath}
= \int d\log_{10}\tau\int d\log_{10}\mass
\end{displaymath}
\begin{equation}
P(\log_{10}\tau,\log_{10}\mass,A_G,T_\textrm{eff},\log g,[M/H],A_0,A_G,d,R,M_G|\vec s,G,\varpi)
.\end{equation}
We note that while the MCMC sampling itself makes use of log-age, $\log_{10}\tau$, and log-initial mass, $\log_{10}\mass$, these values are not actually provided in \gdr3, which is why they are marginalised out from the user perspective. Nevertheless, these parameters are still necessary for the MCMC in order to establish astrophysically consistent relations, for example~between temperature and radius. Applying Bayes' theorem, we obtain:
\begin{displaymath}
    P(T_\textrm{eff},\log g,[M/H],A_0,A_G,d,R,M_G|\vec s,G,\varpi)
\end{displaymath}
\begin{displaymath}
\propto \int d\log_{10}\tau\int d\log_{10}\mass
\end{displaymath}
\begin{displaymath}\phantom{\propto}
P(\vec s,G,\varpi|\log_{10}\tau,\log_{10}\mass,A_G,T_\textrm{eff},\log g,[M/H],A_0,d,R,M_G)
\end{displaymath}
\begin{displaymath}\phantom{\propto}
\cdot P(\log_{10}\tau,\log_{10}\mass,A_G,T_\textrm{eff},\log g,[M/H],A_0,d,R,M_G)
\end{displaymath}
\begin{displaymath}
= \int d\log_{10}\tau\int d\log_{10}\mass
\end{displaymath}
\begin{displaymath}\phantom{\propto}
P(\vec s,G,\varpi|\log_{10}\tau,\log_{10}\mass,A_G,T_\textrm{eff},\log g,[M/H],A_0,d,R,M_G)
\end{displaymath}
\begin{displaymath}\phantom{\propto}
\cdot P(A_G|\log_{10}\tau,\log_{10}\mass,T_\textrm{eff},\log g,[M/H],A_0,d,R,M_G)
\end{displaymath}
\begin{displaymath}\phantom{\propto}
\cdot P(A_0|\log_{10}\tau,\log_{10}\mass,T_\textrm{eff},\log g,[M/H],d,R,M_G)
\end{displaymath}
\begin{displaymath}\phantom{\propto}
\cdot P(d|\log_{10}\tau,\log_{10}\mass,T_\textrm{eff},\log g,[M/H],R,M_G)
\end{displaymath}
\begin{displaymath}\phantom{\propto}
\cdot P(R|\log_{10}\tau,\log_{10}\mass,[M/H],T_\textrm{eff},\log g,M_G)
\end{displaymath}
\begin{displaymath}\phantom{\propto}
\cdot P(\log_{10}\tau,\log_{10}\mass,[M/H],T_\textrm{eff},\log g,M_G).
\end{displaymath}
The first factor is the likelihood of the observables. We can safely assume that the observables $\vec s,G,$ and $ \varpi$ are statistically independent measurements by the Gaia satellite, such that their likelihoods factorise. Dropping all irrelevant dependencies, we therefore obtain:
\begin{displaymath}
    P(T_\textrm{eff},\log g,[M/H],A_0,A_G,d,R,M_G|\vec s,G,\varpi)
\end{displaymath}
\begin{displaymath}
\propto  \int d\log_{10}\tau\int d\log_{10}\mass
\end{displaymath}
\begin{displaymath}\phantom{\propto}
P(\vec s|T_\textrm{eff},\log g,[M/H],A_0,d,R)
\cdot P(G|A_G,d,M_G)
\cdot P(\varpi|d)
\end{displaymath}
\begin{displaymath}\phantom{\propto}
\cdot P(A_G|T_\textrm{eff},\log g,[M/H],A_0)
\end{displaymath}
\begin{displaymath}\phantom{\propto}
\cdot P(A_0|d)
\end{displaymath}
\begin{displaymath}\phantom{\propto}
\cdot P(d)
\end{displaymath}
\begin{displaymath}\phantom{\propto}
\cdot P(R|\log_{10}\tau,\log_{10}\mass,[M/H])
\end{displaymath}
\begin{equation}
\label{eq:appendix-posterior}
\phantom{\propto}
\cdot P([M/H],T_\textrm{eff},\log g,M_G,\log_{10}\tau,\log_{10}\mass).
\end{equation}
This is the fully simplified posterior that is optimised by the Aeneas MCMC.

\subsection{MCMC configuration}
\label{appendix:mcmc-configuration}

As explained in Sect.~\ref{sect:computational-cost}, \gspphot\ makes use of the \emcee\ algorithm \citep{EMCEE2013}. For the ensemble size, we choose 100 walkers in order to explore the 4D parameter space of age, initial mass, metallicity, and $A_0$. We then set up a procedure which we find minimises the risk of the \emcee\ getting stuck in the next best local optimum: First, we initialise the \emcee\ ensemble in a small ball around the initial guess and let it expand for 50 iterations. After these initial 50 iterations we repeat the following procedure five times:
\begin{enumerate}
\item From all previous samples (not only the last ensemble state), identify the 100 best samples, having the highest posterior probability (without repetition of samples).
\item Re-initialise the EMCEE ensemble with these 100 best walkers. Erase previous \emcee\ history.
\item Run for 25 iterations.
\end{enumerate}
After this procedure, we assume that the \emcee\ ensemble has converged and we run it for another 145 iterations. From this final phase, we start from the last ensemble state (100 samples) and work backwards through the MCMC chain taking an ensemble snapshot every 7th iteration until we have gathered a total of 2000 samples. These 2000 samples are then used to estimate the reported median values and confidence intervals.

We note that a thin-out factor of 7 is most likely insufficient to guarantee absence of autocorrelations in the samples. Likewise, after the fifth and last clipping of the \emcee\ ensemble we only have 36 iterations before taking the first ensemble snapshot for inference, which is not always sufficient to guarantee relaxation. As explained in Sect.~\ref{sect:computational-cost}, these choices are the results of limited computational resources. Experiments with longer MCMC chains and more ensemble walkers only show a mild improvement of scientific results, i.e.~this is no major limitation.

\section{Example ADQL queries}
\label{appendix:example-ADQL-queries}

The following query produces the random sample used in Fig.~\ref{fig:CMD-de-reddening} and Fig.~\ref{fig:HRD-and-Kiel}.

\begin{verbatim}
SELECT
gaia.source_id,
gaia.parallax,
gaia.parallax_error,
gaia.phot_g_mean_mag,
gaia.phot_bp_mean_mag,
gaia.phot_rp_mean_mag,
gaia.teff_gspphot,
gaia.logg_gspphot,
gaia.ebpminrp_gspphot,
apsis.mg_gspphot
FROM (
SELECT
source_id,parallax,parallax_error,
phot_g_mean_mag,phot_bp_mean_mag,phot_rp_mean_mag,
teff_gspphot,logg_gspphot,ebpminrp_gspphot
FROM user_dr3int5.gaia_source
WHERE random_index<10000000 
AND teff_gspphot IS NOT NULL
) AS gaia
JOIN user_dr3int5.astrophysical_parameters AS apsis
ON gaia.source_id=apsis.source_id
\end{verbatim}

\noindent
The query below produces the sample of the Local Bubble used in Fig.~\ref{fig:local-bubble-extinction}.

\begin{verbatim}
SELECT
gaia.source_id,
gaia.parallax,
gaia.parallax_error,
gaia.phot_g_mean_mag,
gaia.phot_bp_mean_mag,
gaia.phot_rp_mean_mag,
apsis.azero_gspphot,
apsis.ag_gspphot,
apsis.abp_gspphot,
apsis.arp_gspphot
FROM (
    SELECT
    source_id,parallax,parallax_error,
    phot_g_mean_mag,
    phot_bp_mean_mag,phot_rp_mean_mag
    FROM user_dr3int5.gaia_source
    WHERE parallax>20 AND teff_gspphot IS NOT NULL
) AS gaia
JOIN user_dr3int5.astrophysical_parameters AS apsis
ON gaia.source_id=apsis.source_id
\end{verbatim}

\section{Acknowledgements}\label{ssec:appendixAcknowledgements}

We thank our DPAC colleagues from CU5, Paolo Montegriffo, Dafydd Wyn Evans, Michael Weiler, Carme Jordi, Elena Pancino and Carla Cacciari, who have continuously supported us with their expertise on BP/RP spectra, their instrument characteristics and calibration. We also thank our DPAC colleagues from CU9, Carine Babusiaux, Merc\`e Romero-G\'omez and Francesca Figueras, for their validation work and valuable feedback. Last but not least, we thank our former colleagues Tri Astraadmadja, Dae-Won Kim, Kester Smith, Paravskevi Tsalmantza, Rainer Klement, and Carola Tiede.
%
%This work was funded in part by the DLR (German space agency) via grants 50QG0602, 50QG1001, 50QG1403, and 50QG201.
%It has made use of data from the European Space Agency (ESA) mission Gaia (\url{http://www.cosmos.esa.int/gaia}), processed by the Gaia Data Processing and Analysis Consortium (DPAC, \url{http://www.cosmos.esa.int/web/gaia/dpac/consortium}). Funding for the DPAC has been provided by national institutions, in particular the institutions participating in the Gaia Multilateral Agreement. 
%
This research was achieved using the POLLUX database (\url{http://pollux.oreme.org}) operated at LUPM  (Université Montpellier - CNRS, France with the support of the PNPS and INSU).
% If deemed appropriate, if not kill:
%AJK acknowledges support by the Swedish National Space Agency (SNSA).

This work presents results from the European Space Agency (ESA) space mission \gaia. \gaia\ data are being processed by the \gaia\ Data Processing and Analysis Consortium (DPAC). Funding for the DPAC is provided by national institutions, in particular the institutions participating in the \gaia\ MultiLateral Agreement (MLA). The \gaia\ mission website is \url{https://www.cosmos.esa.int/gaia}. The \gaia\ archive website is \url{https://archives.esac.esa.int/gaia}.

The \gaia\ mission and data processing have financially been supported by, in alphabetical order by country:
\begin{itemize}
\item the Algerian Centre de Recherche en Astronomie, Astrophysique et G\'{e}ophysique of Bouzareah Observatory;
\item the Austrian Fonds zur F\"{o}rderung der wissenschaftlichen Forschung (FWF) Hertha Firnberg Programme through grants T359, P20046, and P23737;
\item the BELgian federal Science Policy Office (BELSPO) through various PROgramme de D\'{e}veloppement d'Exp\'{e}riences scientifiques (PRODEX) grants and the Polish Academy of Sciences - Fonds Wetenschappelijk Onderzoek through grant VS.091.16N, and the Fonds de la Recherche Scientifique (FNRS), and the Research Council of Katholieke Universiteit (KU) Leuven through grant C16/18/005 (Pushing AsteRoseismology to the next level with TESS, GaiA, and the Sloan DIgital Sky SurvEy -- PARADISE);  
\item the Brazil-France exchange programmes Funda\c{c}\~{a}o de Amparo \`{a} Pesquisa do Estado de S\~{a}o Paulo (FAPESP) and Coordena\c{c}\~{a}o de Aperfeicoamento de Pessoal de N\'{\i}vel Superior (CAPES) - Comit\'{e} Fran\c{c}ais d'Evaluation de la Coop\'{e}ration Universitaire et Scientifique avec le Br\'{e}sil (COFECUB);
\item the Chilean Agencia Nacional de Investigaci\'{o}n y Desarrollo (ANID) through Fondo Nacional de Desarrollo Cient\'{\i}fico y Tecnol\'{o}gico (FONDECYT) Regular Project 1210992 (L.~Chemin);
\item the National Natural Science Foundation of China (NSFC) through grants 11573054, 11703065, and 12173069, the China Scholarship Council through grant 201806040200, and the Natural Science Foundation of Shanghai through grant 21ZR1474100;  
\item the Tenure Track Pilot Programme of the Croatian Science Foundation and the \'{E}cole Polytechnique F\'{e}d\'{e}rale de Lausanne and the project TTP-2018-07-1171 `Mining the Variable Sky', with the funds of the Croatian-Swiss Research Programme;
\item the Czech-Republic Ministry of Education, Youth, and Sports through grant LG 15010 and INTER-EXCELLENCE grant LTAUSA18093, and the Czech Space Office through ESA PECS contract 98058;
\item the Danish Ministry of Science;
\item the Estonian Ministry of Education and Research through grant IUT40-1;
\item the European Commission’s Sixth Framework Programme through the European Leadership in Space Astrometry (\href{https://www.cosmos.esa.int/web/gaia/elsa-rtn-programme}{ELSA}) Marie Curie Research Training Network (MRTN-CT-2006-033481), through Marie Curie project PIOF-GA-2009-255267 (Space AsteroSeismology \& RR Lyrae stars, SAS-RRL), and through a Marie Curie Transfer-of-Knowledge (ToK) fellowship (MTKD-CT-2004-014188); the European Commission's Seventh Framework Programme through grant FP7-606740 (FP7-SPACE-2013-1) for the \gaia\ European Network for Improved data User Services (\href{https://gaia.ub.edu/twiki/do/view/GENIUS/}{GENIUS}) and through grant 264895 for the \gaia\ Research for European Astronomy Training (\href{https://www.cosmos.esa.int/web/gaia/great-programme}{GREAT-ITN}) network;
\item the European Cooperation in Science and Technology (COST) through COST Action CA18104 `Revealing the Milky Way with \gaia (MW-Gaia)';
\item the European Research Council (ERC) through grants 320360, 647208, and 834148 and through the European Union’s Horizon 2020 research and innovation and excellent science programmes through Marie Sk{\l}odowska-Curie grant 745617 (Our Galaxy at full HD -- Gal-HD) and 895174 (The build-up and fate of self-gravitating systems in the Universe) as well as grants 687378 (Small Bodies: Near and Far), 682115 (Using the Magellanic Clouds to Understand the Interaction of Galaxies), 695099 (A sub-percent distance scale from binaries and Cepheids -- CepBin), 716155 (Structured ACCREtion Disks -- SACCRED), 951549 (Sub-percent calibration of the extragalactic distance scale in the era of big surveys -- UniverScale), and 101004214 (Innovative Scientific Data Exploration and Exploitation Applications for Space Sciences -- EXPLORE);
\item the European Science Foundation (ESF), in the framework of the \gaia\ Research for European Astronomy Training Research Network Programme (\href{https://www.cosmos.esa.int/web/gaia/great-programme}{GREAT-ESF});
\item the European Space Agency (ESA) in the framework of the \gaia\ project, through the Plan for European Cooperating States (PECS) programme through contracts C98090 and 4000106398/12/NL/KML for Hungary, through contract 4000115263/15/NL/IB for Germany, and through PROgramme de D\'{e}veloppement d'Exp\'{e}riences scientifiques (PRODEX) grant 4000127986 for Slovenia;  
\item the Academy of Finland through grants 299543, 307157, 325805, 328654, 336546, and 345115 and the Magnus Ehrnrooth Foundation;
\item the French Centre National d’\'{E}tudes Spatiales (CNES), the Agence Nationale de la Recherche (ANR) through grant ANR-10-IDEX-0001-02 for the `Investissements d'avenir' programme, through grant ANR-15-CE31-0007 for project `Modelling the Milky Way in the \gaia era’ (MOD4Gaia), through grant ANR-14-CE33-0014-01 for project `The Milky Way disc formation in the \gaia era’ (ARCHEOGAL), through grant ANR-15-CE31-0012-01 for project `Unlocking the potential of Cepheids as primary distance calibrators’ (UnlockCepheids), through grant ANR-19-CE31-0017 for project `Secular evolution of galxies' (SEGAL), and through grant ANR-18-CE31-0006 for project `Galactic Dark Matter' (GaDaMa), the Centre National de la Recherche Scientifique (CNRS) and its SNO \gaia of the Institut des Sciences de l’Univers (INSU), its Programmes Nationaux: Cosmologie et Galaxies (PNCG), Gravitation R\'{e}f\'{e}rences Astronomie M\'{e}trologie (PNGRAM), Plan\'{e}tologie (PNP), Physique et Chimie du Milieu Interstellaire (PCMI), and Physique Stellaire (PNPS), the `Action F\'{e}d\'{e}ratrice \gaia' of the Observatoire de Paris, the R\'{e}gion de Franche-Comt\'{e}, the Institut National Polytechnique (INP) and the Institut National de Physique nucl\'{e}aire et de Physique des Particules (IN2P3) co-funded by CNES;
\item the German Aerospace Agency (Deutsches Zentrum f\"{u}r Luft- und Raumfahrt e.V., DLR) through grants 50QG0501, 50QG0601, 50QG0602, 50QG0701, 50QG0901, 50QG1001, 50QG1101, 50\-QG1401, 50QG1402, 50QG1403, 50QG1404, 50QG1904, 50QG2101, 50QG2102, and 50QG2202, and the Centre for Information Services and High Performance Computing (ZIH) at the Technische Universit\"{a}t Dresden for generous allocations of computer time;
\item the Hungarian Academy of Sciences through the Lend\"{u}let Programme grants LP2014-17 and LP2018-7 and the Hungarian National Research, Development, and Innovation Office (NKFIH) through grant KKP-137523 (`SeismoLab');
\item the Science Foundation Ireland (SFI) through a Royal Society - SFI University Research Fellowship (M.~Fraser);
\item the Israel Ministry of Science and Technology through grant 3-18143 and the Tel Aviv University Center for Artificial Intelligence and Data Science (TAD) through a grant;
\item the Agenzia Spaziale Italiana (ASI) through contracts I/037/08/0, I/058/10/0, 2014-025-R.0, 2014-025-R.1.2015, and 2018-24-HH.0 to the Italian Istituto Nazionale di Astrofisica (INAF), contract 2014-049-R.0/1/2 to INAF for the Space Science Data Centre (SSDC, formerly known as the ASI Science Data Center, ASDC), contracts I/008/10/0, 2013/030/I.0, 2013-030-I.0.1-2015, and 2016-17-I.0 to the Aerospace Logistics Technology Engineering Company (ALTEC S.p.A.), INAF, and the Italian Ministry of Education, University, and Research (Ministero dell'Istruzione, dell'Universit\`{a} e della Ricerca) through the Premiale project `MIning The Cosmos Big Data and Innovative Italian Technology for Frontier Astrophysics and Cosmology' (MITiC);
\item the Netherlands Organisation for Scientific Research (NWO) through grant NWO-M-614.061.414, through a VICI grant (A.~Helmi), and through a Spinoza prize (A.~Helmi), and the Netherlands Research School for Astronomy (NOVA);
\item the Polish National Science Centre through HARMONIA grant 2018/30/M/ST9/00311 and DAINA grant 2017/27/L/ST9/03221 and the Ministry of Science and Higher Education (MNiSW) through grant DIR/WK/2018/12;
\item the Portuguese Funda\c{c}\~{a}o para a Ci\^{e}ncia e a Tecnologia (FCT) through national funds, grants SFRH/\-BD/128840/2017 and PTDC/FIS-AST/30389/2017, and work contract DL 57/2016/CP1364/CT0006, the Fundo Europeu de Desenvolvimento Regional (FEDER) through grant POCI-01-0145-FEDER-030389 and its Programa Operacional Competitividade e Internacionaliza\c{c}\~{a}o (COMPETE2020) through grants UIDB/04434/2020 and UIDP/04434/2020, and the Strategic Programme UIDB/\-00099/2020 for the Centro de Astrof\'{\i}sica e Gravita\c{c}\~{a}o (CENTRA);  
\item the Slovenian Research Agency through grant P1-0188;
\item the Spanish Ministry of Economy (MINECO/FEDER, UE), the Spanish Ministry of Science and Innovation (MICIN), the Spanish Ministry of Education, Culture, and Sports, and the Spanish Government through grants BES-2016-078499, BES-2017-083126, BES-C-2017-0085, ESP2016-80079-C2-1-R, ESP2016-80079-C2-2-R, FPU16/03827, PDC2021-121059-C22, RTI2018-095076-B-C22, and TIN2015-65316-P (`Computaci\'{o}n de Altas Prestaciones VII'), the Juan de la Cierva Incorporaci\'{o}n Programme (FJCI-2015-2671 and IJC2019-04862-I for F.~Anders), the Severo Ochoa Centre of Excellence Programme (SEV2015-0493), and MICIN/AEI/10.13039/501100011033 (and the European Union through European Regional Development Fund `A way of making Europe') through grant RTI2018-095076-B-C21, the Institute of Cosmos Sciences University of Barcelona (ICCUB, Unidad de Excelencia `Mar\'{\i}a de Maeztu’) through grant CEX2019-000918-M, the University of Barcelona's official doctoral programme for the development of an R+D+i project through an Ajuts de Personal Investigador en Formaci\'{o} (APIF) grant, the Spanish Virtual Observatory through project AyA2017-84089, the Galician Regional Government, Xunta de Galicia, through grants ED431B-2021/36, ED481A-2019/155, and ED481A-2021/296, the Centro de Investigaci\'{o}n en Tecnolog\'{\i}as de la Informaci\'{o}n y las Comunicaciones (CITIC), funded by the Xunta de Galicia and the European Union (European Regional Development Fund -- Galicia 2014-2020 Programme), through grant ED431G-2019/01, the Red Espa\~{n}ola de Supercomputaci\'{o}n (RES) computer resources at MareNostrum, the Barcelona Supercomputing Centre - Centro Nacional de Supercomputaci\'{o}n (BSC-CNS) through activities AECT-2017-2-0002, AECT-2017-3-0006, AECT-2018-1-0017, AECT-2018-2-0013, AECT-2018-3-0011, AECT-2019-1-0010, AECT-2019-2-0014, AECT-2019-3-0003, AECT-2020-1-0004, and DATA-2020-1-0010, the Departament d'Innovaci\'{o}, Universitats i Empresa de la Generalitat de Catalunya through grant 2014-SGR-1051 for project `Models de Programaci\'{o} i Entorns d'Execuci\'{o} Parallels' (MPEXPAR), and Ramon y Cajal Fellowship RYC2018-025968-I funded by MICIN/AEI/10.13039/501100011033 and the European Science Foundation (`Investing in your future');
\item the Swedish National Space Agency (SNSA/Rymdstyrelsen);
\item the Swiss State Secretariat for Education, Research, and Innovation through the Swiss Activit\'{e}s Nationales Compl\'{e}mentaires and the Swiss National Science Foundation through an Eccellenza Professorial Fellowship (award PCEFP2\_194638 for R.~Anderson);
\item the United Kingdom Particle Physics and Astronomy Research Council (PPARC), the United Kingdom Science and Technology Facilities Council (STFC), and the United Kingdom Space Agency (UKSA) through the following grants to the University of Bristol, the University of Cambridge, the University of Edinburgh, the University of Leicester, the Mullard Space Sciences Laboratory of University College London, and the United Kingdom Rutherford Appleton Laboratory (RAL): PP/D006511/1, PP/D006546/1, PP/D006570/1, ST/I000852/1, ST/J005045/1, ST/K00056X/1, ST/\-K000209/1, ST/K000756/1, ST/L006561/1, ST/N000595/1, ST/N000641/1, ST/N000978/1, ST/\-N001117/1, ST/S000089/1, ST/S000976/1, ST/S000984/1, ST/S001123/1, ST/S001948/1, ST/\-S001980/1, ST/S002103/1, ST/V000969/1, ST/W002469/1, ST/W002493/1, ST/W002671/1, ST/W002809/1, and EP/V520342/1.
\end{itemize}

The GBOT programme  uses observations collected at (i) the European Organisation for Astronomical Research in the Southern Hemisphere (ESO) with the VLT Survey Telescope (VST), under ESO programmes
092.B-0165,
093.B-0236,
094.B-0181,
095.B-0046,
096.B-0162,
097.B-0304,
098.B-0030,
099.B-0034,
0100.B-0131,
0101.B-0156,
0102.B-0174, and
0103.B-0165;
%
% From Martin Altmann, 13 March 2019:
%  092.B-0165   01.10.13 - 31.03.14
%  093.B-0236   01.04.14 - 30.09.14
%  094.B-0181   01.10.14 - 31.03.15
%  095.B-0046   01.04.15 - 30.09.15
%  096.B-0162   01.10.15 - 31.03.16
%  097.B-0304   01.04.16 - 30.09.16
%  098.B-0030   01.10.16 - 31.03.17
%  099.B-0034   01.04.17 - 30.09.17
% 0100.B-0131   01.10.17 - 31.03.18
% 0101.B-0156   01.04.18 - 30.09.18
% 0102.B-0174   01.10.18 - 31.03.19
% 0103.B-0165   01.04.19 - 30.09.19
%
and (ii) the Liverpool Telescope, which is operated on the island of La Palma by Liverpool John Moores University in the Spanish Observatorio del Roque de los Muchachos of the Instituto de Astrof\'{\i}sica de Canarias with financial support from the United Kingdom Science and Technology Facilities Council, and (iii) telescopes of the Las Cumbres Observatory Global Telescope Network.

\end{document}